\documentclass[pra,twocolumn,superscriptaddress]{revtex4-1}

\usepackage[utf8]{inputenc}
\usepackage[english]{babel}
\usepackage[T1]{fontenc}
\usepackage{lmodern}

\usepackage{graphicx}

\usepackage{amsmath}
\usepackage{amsfonts}
\usepackage{amssymb}
\usepackage{amsthm}

\usepackage{mathtools}

\usepackage{dsfont}
\usepackage{microtype}

\usepackage[breaklinks=true,colorlinks]{hyperref}

\newcommand{\cf}{cf.\ }

\newcommand{\coloneq}{\mathrel{\mathop:}=}
\newcommand{\dd}{\mathrm{d}}
\newcommand{\Tr}{\operatorname{Tr}}
\newcommand{\bkew}[3]{\left\langle{#1}\middle|{#2}\middle|{#3}\right\rangle}
\newcommand{\bkewabs}[3]{\left|\left\langle{#1}\middle|{#2}\middle|{#3}\right\rangle\right|}
\newcommand{\ket}[1]{\left|{#1}\right\rangle}

\newcommand{\ketbra}[2]{\left|{#1}\middle\rangle\middle\langle{#2}\right|}
\newcommand{\ew}[1]{\left\langle{#1}\right\rangle}

\newcommand{\kB}{k_\mathrm{B}}
\newcommand{\indexc}{\mathrm{c}}
\newcommand{\indexh}{\mathrm{h}}
\newcommand{\betaeff}{\beta_\mathrm{eff}}
\newcommand{\betac}{\beta_\indexc}
\newcommand{\betah}{\beta_\indexh}
\newcommand{\Tc}{T_\indexc}
\newcommand{\Th}{T_\indexh}
\newcommand{\Jc}{\mathcal{J}_\indexc}

\newcommand{\nbar}{\bar{n}}
\newcommand{\J}{\mathfrak{j}}

\begin{document}

\title{Cooperative many-body enhancement of quantum thermal machine power}

\author{Wolfgang Niedenzu}
\email{Wolfgang.Niedenzu@uibk.ac.at}
\affiliation{Institut f\"ur Theoretische Physik, Universit\"at Innsbruck, Technikerstra{\ss}e~21a, A-6020~Innsbruck, Austria}

\author{Gershon Kurizki}
\affiliation{Department of Chemical Physics, Weizmann Institute of Science, Rehovot~7610001, Israel}

\begin{abstract}
  We study the impact of cooperative many-body effects on the operation of periodically-driven quantum thermal machines, particularly heat engines and refrigerators. In suitable geometries, $N$ two-level atoms can exchange energy with the driving field and the (hot and cold) thermal baths at a faster rate than a single atom due to their SU(2) symmetry that causes the atoms to behave as a collective spin-$N/2$ particle. This cooperative effect boosts the power output of heat engines compared to the power output of $N$ independent, incoherent, heat engines. In the refrigeration regime, similar cooling-power boost takes place.
\end{abstract}

\date{November 2, 2018}

\maketitle

\section{Introduction}

One of the central questions in the emerging field of quantum thermodynamics~\cite{kosloff2013quantum,gelbwaser2015thermodynamics,goold2016role,vinjanampathy2016quantum,alicki2018introduction} pertains to possible quantum advantages (``supremacy'') in the operation of thermal machines such as heat engines or refrigerators compared to their classical counterparts~\cite{cengelbook}. In that context, starting with the seminal work by Scully et al.~\cite{scully2003extracting}, extensive investigations have focused on the question whether quantum coherence in either the machine's working medium~\cite{scully2011quantum,brandner2015coherence,uzdin2015equivalence,niedenzu2015performance,uzdin2016coherence,friedenberger2017quantum} or the energising (hot) bath (the ``fuel'')~\cite{dillenschneider2009energetics,huang2012effects,abah2014efficiency,rossnagel2014nanoscale,hardal2015superradiant,dag2016multiatom,manzano2016entropy,turkpence2016quantum,agarwalla2017quantum,dag2018temperature,niedenzu2018quantum} could either boost the power output or the efficiency of quantum engines. Whilst these investigations have been mainly theoretical, impressive experimental progress has also been made such as the first realisation of a heat engine based on a single atom~\cite{rossnagel2016single}, the demonstration of quantum-thermodynamic effects in the operation of a heat engine implemented by an ensemble of nitrogen-vacancy (NV) centres in diamond~\cite{klatzow2017experimental} and the simulation of a quantum engine fuelled by a squeezed-thermal bath in a classical setting~\cite{klaers2017squeezed}.

\par

Here we explore the possibility of exploiting collective (cooperative) many-body effects in quantum heat engines and refrigerators~\cite{jiang2014enhancing,binder2015quantacell,campisi2016power,jaramillo2016quantum,campaioli2017enhancing,vroylandt2017collective,ferraro2018high,hardal2018phase,le2018spin}. These generic quantum effects have a common origin with Dicke superradiance~\cite{dicke1954coherence}, whereby light emission is collectively enhanced by the interaction of $N$ atoms with a common environment (bath) such that its intensity scales with $N^2$~\cite{dicke1954coherence,agarwal1970master,lehmberg1970radiation,lehmberg1970radiation2,skribanowitz1973observation,carmichael1980analytical,gross1982superradiance,devoe1996observation,hald1999spin,wang2007superradiance,akkermans2008photon,chang2013self,ritsch2013cold,vanloo2013photon,wickenbrock2013collective,maier2014superradiance,meir2014cooperative,mlynek2014observation,zou2014implementation,goban2015superradiance,klinder2015observation,araujo2016superradiance,shammah2017superradiance,angerer2018superradiant,mandelbook,breuerbook}. Investigations of this effect for a cloud of closely-packed emitters~\cite{gross1982superradiance} face a severe problem: The dipole-dipole interaction (DDI) among emitters may diverge and cause an uncontrollable inhomogeneous broadening that may destroy superradiance. By contrast, DDI may be either suppressed~\cite{vanloo2013photon} or collectively enhanced~\cite{shahmoon2013nonradiative} in appropriate one-dimensional setups~\cite{lalumiere2013input} such that in either case superradiance persists.

\par

Beyond the need to control DDI, the feasibility of quantum cooperative effects in heat machines depends on the resolution of another principal issue: Is the timing of atom-bath interactions important for their cooperativity~\cite{scully2006directed,mazets2007multiatom,svidzinsky2008dynamical}? Since the early studies of superradiance and superfluorescence~\cite{bonifacio1975cooperative} the crucial role of proper initiation of the atomic ensemble has been stressed~\cite{scully2009super,svidzinsky2008fermi,svidzinsky2010cooperative}. If this is the case, is \emph{continuous}, steady-state interaction of the atoms with heat baths compatible with cooperativity? Here we show that, rather surprisingly, quantum cooperative enhancement of $N$-atom energy exchange with thermal baths persists under steady-state conditions. This fundamental result is an outcome of our analysis that extends to $N$ atoms the concept of the minimal quantum heat machine~\cite{gelbwaser2013minimal,alicki2014quantum}: A two-level atom that is simultaneously coupled to two (hot and cold) thermal baths and is periodically modulated in energy (transition frequency).

\par

This paper is organised as follows: We introduce the model for an $N$-partite quantum thermal machine and discuss the collective-spin basis (Dicke basis) in Sec.~\ref{sec_model}. The thermodynamic analysis is based on the Markovian master equation presented in Sec.~\ref{sec_master}. The collective energy currents and the power boost compared to $N$ independent engines are analysed in Secs.~\ref{sec_currents} and~\ref{sec_power}. A concrete example for a heat engine and a refrigerator is discussed in Secs.~\ref{sec_sin} and~\ref{sec_refrigerator}. In Sec.~\ref{sec_realisation} we address possible experimental realisations before concluding in Sec.~\ref{sec_conclusions}.

\section{Many-body thermal machines: model and collective basis}\label{sec_model}

We consider an ensemble of $N$ two-level atoms subject to a periodic driving field that modulates the atomic transition frequency~\cite{gelbwaser2013minimal,alicki2014quantum,gelbwaser2015thermodynamics},
\begin{subequations}\label{eq_H}
  \begin{equation}\label{eq_H_S}
    H_\mathrm{S}(t)=\frac{\hbar\omega(t)}{2}\sum_{k=1}^N\sigma_z^k,
  \end{equation}
  such that $H_\mathrm{S}(t+\tau)=H_\mathrm{S}(t)$. The modulation fulfills the periodicity condition $\tau^{-1}\int_0^\tau\omega(t)\dd t=\omega_0$, where $\tau=2\pi/\Omega$ is the cycle time and $\omega_0$ the atoms' ``bare'' (unperturbed) transition frequency. The atoms are all \emph{identically} coupled to two (cold and hot) thermal baths ($i\in\{\indexc,\indexh\}$),
  \begin{equation}\label{eq_H_SB}
    H_\mathrm{SB}^i=\sum_{k=1}^N\sigma_x^k\otimes B_i,
  \end{equation}
\end{subequations}
where $B_i$ is the coupling operator of the $i$th bath~\cite{breuerbook,wallsbook,carmichaelbook}. Note that the atoms are indistinguishable to both of the baths.

\par
\begin{figure}
  \centering
  \includegraphics[width=0.7\columnwidth]{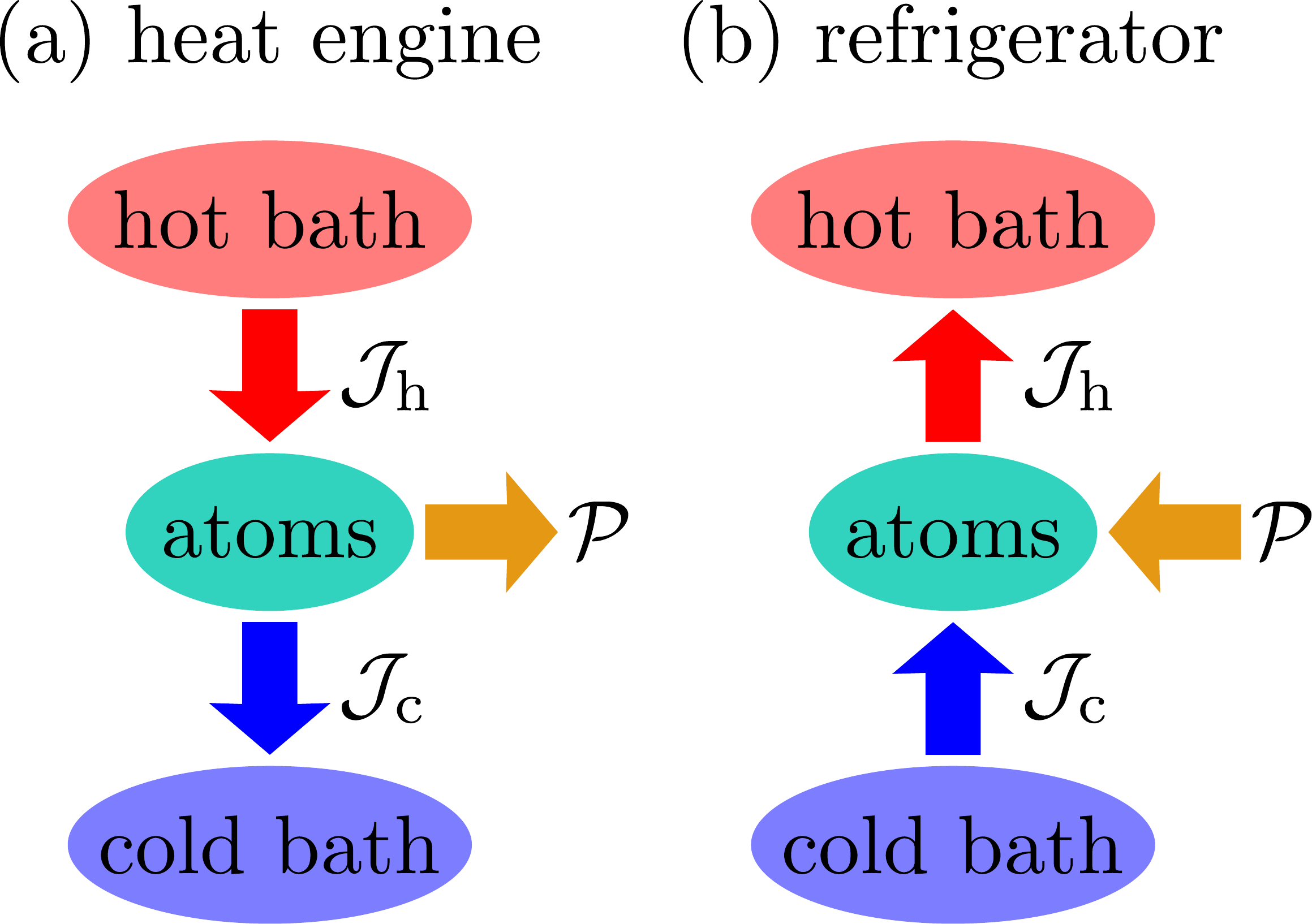}
  \caption{Energy currents in (a)~a heat engine and (b)~a refrigerator. A heat engine converts the heat current $\mathcal{J}_\indexh$ from the hot bath into power $\mathcal{P}$ whereas a refrigerator consumes power to establish a heat flow from the cold to the hot bath. Regardless of the operation mode, we here show that all three energy currents $\mathcal{J}_\indexc$, $\mathcal{J}_\indexh$ and $\mathcal{P}$ can be equally enhanced by cooperative effects between the atoms. We use the convention that energy currents directed towards the atoms have positive sign. Heat-engine operation then corresponds to $\mathcal{P}<0$.}\label{fig_thermal_machines}
\end{figure}

\par

The thermodynamic properties of a setup governed by Eqs.~\eqref{eq_H} have been studied for $N=1$, i.e., a single two-level atom, in Refs.~\cite{gelbwaser2013minimal,alicki2014quantum}. In particular, it has been shown that this setup constitutes a universal thermal machine that can be operated on-demand as either a heat engine (thereby converting heat obtained from the hot bath into power) or a refrigerator (consuming power to cool the cold bath), see Fig.~\ref{fig_thermal_machines}. The classical field that induces the periodic modulation in the Hamiltonian~\eqref{eq_H_S} acts as a piston that either extracts (supplies) power from (to) the machine~\cite{kosloff2013quantum}.

\par

We now introduce the collective spin operator $\mathbf{J}\coloneq(J_x,J_y,J_z)^T$ with components~\cite{breuerbook}
\begin{subequations}\label{eq_J}
  \begin{equation}
    J_j\coloneq\sum_{k=1}^N\frac{1}{2}\sigma_j^k \quad (j\in\{x,y,z\})
  \end{equation}
  and the transition operators
  \begin{equation}\label{eq_J_pm}
    J_\pm\coloneq\sum_{k=1}^N\sigma_\pm^k.
  \end{equation}
\end{subequations}
In terms of these collective operators, the Hamiltonians~\eqref{eq_H} adopt the simple forms
\begin{subequations}\label{eq_H_J}
  \begin{align}
    H_\mathrm{S}(t)&=\hbar\omega(t) J_z\label{eq_H_J_HS}\\
    H_\mathrm{SB}^i&=2J_x\otimes B_i\label{eq_H_J_HSB}.
  \end{align}
\end{subequations}
The pertinent feature of the collective operators~\eqref{eq_J} is that they can be cast into the block-diagonal form
\begin{equation}\label{eq_J_Sk}
  J_j=\bigoplus_{k=1}^n S_j^k\quad(j\in\{x,y,z,\pm\}),
\end{equation}
where the $S_j^k$ are spin operators of lower dimension. The block-diagonal form~\eqref{eq_J_Sk} is adopted in the basis spanned by the Dicke states, which are entangled many-body states~\cite{mandelbook,breuerbook,tasgin2017many}.

\par

The decomposition~\eqref{eq_J_Sk} follows from the theory of spin addition~\cite{schwablbookqm1,breuerbook}: The $2^N$-dimensional joint Hilbert space of $N$ two-level atoms can be decomposed into $n$ irreducible subspaces of dimension $2\J+1$, each corresponding to an eigenvalue $\J(\J+1)$ of $\mathbf{J}^2$~\cite{breuerbook}. The possible values of $\J$ are $\J=0,1,\dots,\frac{N}{2}$ for $N$ even and $\J=\frac{1}{2},\frac{3}{2},\dots,\frac{N}{2}$ for $N$ odd, respectively. We will make use of the notation $\J_k$ to denote the $\J$ belonging to the $k$th subspace ($k=1,\dots,n$) when necessary.

\par

As an example, two spin-$1/2$ particles couple to a triplet ($\J=1$) and a singlet ($\J=0$), whereas three spin-$1/2$ couple to a quadruplet ($\J=\frac{3}{2}$) and two doublets ($\J=\frac{1}{2}$). These decompositions~\cite{zachos1992altering} are commonly denoted as $2\otimes2=3\oplus1$ and $2\otimes2\otimes2=4\oplus2\oplus2$, respectively, where the numbers indicate the dimensionality $2\J+1$ of the Hilbert space of the respective spin. The latter example shows that different irreducible subspaces may share the same $\J$ (called multiplicity of $\J$). The block-diagonal structure of the collective operators is nevertheless still guaranteed by the eigenstates' different symmetry~\cite{mandelbook}. The largest possible spin $\J=N/2$, however, is unique and formed by the totally-symmetric $N$-atom states (see \ref{app_spin_N_half})~\cite{breuerbook}.

\par
\begin{figure}
  \centering
  \includegraphics[width=0.9\columnwidth]{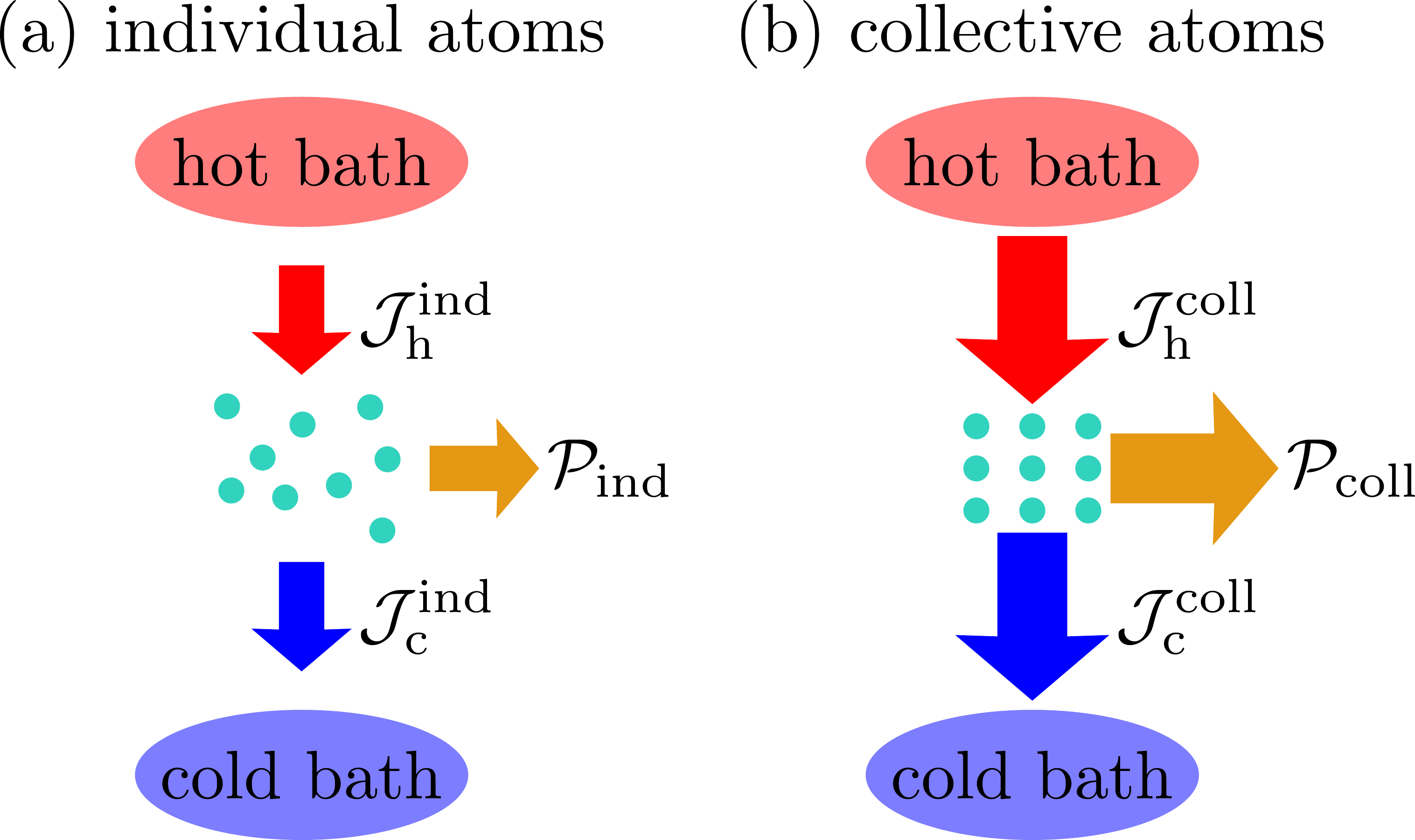}
  \caption{Illustration of the collective boost of the energy currents in the heat engine from Fig.~\ref{fig_thermal_machines}a. (a)~A heat engine comprised of $N$ individual two-level atoms (without any direct or indirect interactions) positions behaves like $N$ individual engines where the energy currents are additive and hence scale linearly in $N$. (b)~In certain geometries where the atoms are ordered and appear indistinguishable to the baths the energy currents can be collectively enhanced owing to constructive interference, $|\mathcal{P}_\mathrm{coll}|>|\mathcal{P}_\mathrm{ind}|$. The same effect also holds for the refrigerator in Fig.~\ref{fig_thermal_machines}b.}\label{fig_heat_engine_boost}
\end{figure}
\par

In this paper we wish to explore the thermodynamic implications of the existence of irreducible subspaces on the operation of thermal machines (see Fig.~\ref{fig_thermal_machines}). Phrased differently, we raise the question: Can the symmetry of the Hamiltonian~\eqref{eq_H_J} be exploited as a genuine quantum-thermodynamic resource that boosts the performance of heat engines or refrigerators? The answer to this question is positive (Fig.~\ref{fig_heat_engine_boost}).

\section{Master equation}\label{sec_master}

A Lindblad master equation for the reduced state of the atoms for periodic Hamiltonians can be derived using Floquet theory~\cite{geva1995relaxation,alicki2006internal,kosloff2013quantum,szczygielski2013markovian,alicki2014quantum,szczygielski2014application,gelbwaser2015thermodynamics,breuerbook}. For our system described by the Hamiltonian~\eqref{eq_H}, the master equation for the single-atom case $N=1$ is given in Refs.~\cite{gelbwaser2013minimal,alicki2014quantum,gelbwaser2015thermodynamics}; for a detailed derivation of the master equation we refer the reader to Sec.~4.1 of Ref.~\cite{alicki2014quantum}. Intriguingly, this derivation also holds for $N>1$ upon replacing the single-atom operators by their collective counterparts according to Eqs.~\eqref{eq_J}; the reason being that $\sigma_j$ and $J_j$ fulfill the same Lie algebra, i.e., the same commutation relations~\cite{schwablbookqm1}. Hence, the multi-atom master equation in the interaction picture reads
\begin{equation}\label{eq_master}
  \dot{\rho}=\sum_{i\in\{\indexc,\indexh\}}\sum_{q\in\mathbb{Z}}\mathcal{L}_{i,q}\rho
\end{equation}
with the sub-Liouvillians
\begin{multline}\label{eq_L}
  \mathcal{L}_{i,q}\rho\coloneq\frac{1}{2}P(q)G_i(\omega_0+q\Omega)\mathcal{D}[J_-] \\
  +\frac{1}{2}P(q)G_i(\omega_0+q\Omega)e^{-\beta_i\hbar(\omega_0+q\Omega)}\mathcal{D}[J_+],
\end{multline}
where $\mathcal{D}[A]\coloneq 2A\rho A^\dagger-A^\dagger A\rho-\rho A^\dagger A$ denotes the dissipator~\cite{breuerbook}.

\par

Physically, the master equation~\eqref{eq_master} describes the incoherent emission (first part of Eq.~\eqref{eq_L} with jump operator $J_-$) and absorption (second part of Eq.~\eqref{eq_L} with jump operator $J_+$) of photons carrying energy $\hbar(\omega_0+q\Omega)$ to (from) the two baths at inverse temperatures $\betac$ and $\betah$, respectively. The frequencies $\omega_0+q\Omega$ ($q\in\mathbb{Z}$) are called Floquet sideband frequencies. The rates associated to these sideband channels are determined by the bath spectra $G_i(\omega)$ and the weights $P(q)$ which depend on the form of the modulation $\omega(t)$ in the Hamiltonian~\eqref{eq_H_S} (see \ref{app_master}). Hence, the master equation~\eqref{eq_master} accounts for the fact that owing to the periodic drive photons emitted to or absorbed from the baths do not necessarily carry the bare two-level transition energy $\hbar\omega_0$ but their energy also depends on the pump frequency $\Omega$. This means that the exchange of photons between the atoms and the baths involves the simultaneous absorption or emission of energy quanta $\hbar\Omega$ from/to the driving field that creates the energy modulation in the Hamiltonian~\eqref{eq_H_S}.

\par

The merit of the collective operators $J_\pm$ [Eq.~\eqref{eq_J_pm}] is now clear: The master equation~\eqref{eq_master} conserves the symmetry, meaning that it does not couple the individual irreducible subspaces that correspond to the blocks in Eq.~\eqref{eq_J_Sk}. In other words, since $[J_\pm,\mathbf{J}^2]=0$ these subspaces are \emph{invariant} under the master equation~\eqref{eq_master} such that every \mbox{$(2\J_k+1)$}-dimensional subspin evolves individually. Hence, the steady-state solution of the master equation~\eqref{eq_master} is a weighted direct sum of Gibbs-like states of these individual spin constituents,
\begin{equation}\label{eq_rho_ss}
  \rho_\mathrm{ss}=\bigoplus_{k=1}^{n}\ew{\Pi_k}_{\rho_0}\rho_\mathrm{ss}^k,
\end{equation}
where
\begin{equation}\label{eq_rho_ss_k}
  \rho_\mathrm{ss}^k=Z_k^{-1}\exp\left(-\betaeff\hbar\omega_0S_z^k\right)
\end{equation}
and $Z_k\coloneq \Tr_k\left[\exp\left(-\betaeff\hbar\omega_0S_z^k\right)\right]$. Here $\Pi_k$ denotes the projector onto the $k$th invariant subspace and $\rho_0$ is the initial condition of the atoms such that $\sum_{k=1}^n\ew{\Pi_k}_{\rho_0}=1$. The inverse effective temperature $\betaeff$ is implicitly defined via the ``global'' detailed-balance condition
\begin{multline}\label{eq_betaeff}
  \exp(-\betaeff\hbar\omega_0)\coloneq\\
  \frac{\sum_{i\in\{\indexc,\indexh\}}\sum_{q\in\mathbb{Z}}P(q)G_i(\omega_0+q\Omega)e^{-\beta_i\hbar(\omega_0+q\Omega)}}{\sum_{i\in\{\indexc,\indexh\}}\sum_{q\in\mathbb{Z}}P(q)G_i(\omega_0+q\Omega)}
\end{multline}
in the master equation~\eqref{eq_master}~\cite{alicki2014quantum}: The nominator in Eq.~\eqref{eq_betaeff} is the total absorption rate in the master equation~\eqref{eq_master} whereas the denominator corresponds to the total emission rate therein.

\par

The steady state~\eqref{eq_rho_ss} thus reflects the fact that the initial populations $\ew{\Pi_k}_{\rho_0}$ of the invariant subspaces (i.e., the initial weights of the individual subspins) cannot change dynamically under the master equation~\eqref{eq_master} since the corresponding subspins do not interact with each other (the effect of imperfections is discussed in~\ref{app_master}). Each subspin, however, relaxes to the Gibbs-like steady state~\eqref{eq_rho_ss_k}.

\par

We note that the static steady state~\eqref{eq_rho_ss} is attained in the interaction picture w.r.t.\ the Hamiltonian~\eqref{eq_H_S}. In the original Schr\"odinger picture this state corresponds to a limit cycle with periodicity $\tau=2\pi/\Omega$~\cite{szczygielski2013markovian,kosloff2013quantum}.

\section{Collective energy currents}\label{sec_currents}

The steady state~\eqref{eq_rho_ss} of the atoms is an out-of-equilibrium state that is maintained by the interplay of three energy currents: (i)~the heat current $\mathcal{J}_\indexc$ from the cold bath, (ii)~the heat current $\mathcal{J}_\indexh$ from the hot bath and (iii)~the power $\mathcal{P}=-[\mathcal{J}_\indexc+\mathcal{J}_\indexh]$ originating from the driving field (see Fig.~\ref{fig_thermal_machines}). Following the theory of the thermodynamics of periodically-driven open quantum systems~\cite{szczygielski2013markovian,kosloff2013quantum,gelbwaser2013minimal,alicki2014quantum,gelbwaser2015thermodynamics}, the heat currents from the two baths to the atoms are ($i\in\{\indexc,\indexh\}$)
\begin{equation}\label{eq_heat}
  \mathcal{J}_i=\sum_{q\in\mathbb{Z}}\hbar(\omega_0+q\Omega)\Tr\left[\left(\mathcal{L}_{i,q}{\rho}_\mathrm{ss}\right)J_z\right].
\end{equation}
This expression accounts for the fact that under $\Omega$-periodic driving photons are not only exchanged at the bare transition frequency~$\omega_0$ but also at the Floquet sidebands $\omega_0+q\Omega$ and thus carry different energies $\hbar(\omega_0+q\Omega)$. These energies are multiplied with the respective steady-state probability for this energy transfer to happen due to the interaction of the spin with bath $i$ at the $q$th harmonic sideband~\cite{alicki2014quantum}.

\par

Inserting the steady state~\eqref{eq_rho_ss} into Eq.~\eqref{eq_heat}, the heat currents and the power evaluate to
\begin{subequations}\label{eq_currents_tot}
  \begin{align}
    \mathcal{J}_i&=\sum_{k=1}^n\ew{\Pi_k}_{\rho_0}\mathcal{J}_i(\J_k)\label{eq_J_tot}\\
    \mathcal{P}&=\sum_{k=1}^n\ew{\Pi_k}_{\rho_0}\mathcal{P}(\J_k)\label{eq_P_tot},
\end{align}
\end{subequations}
where
\begin{multline}\label{eq_J_jk}
  \mathcal{J}_i(\J_k)=F(\J_k)\sum_{q\in\mathbb{Z}}\hbar(\omega_0+q\Omega)P(q)G_i(\omega_0+q\Omega)\\
  \times\left[e^{-\beta_i\hbar(\omega_0+q\Omega)}-e^{-\betaeff\hbar\omega_0}\right]
\end{multline}
is the heat flow induced by the spin-$\J_k$ associated with the $k$th invariant subspace and $\mathcal{P}(\J_k)=-\left[\mathcal{J}_\indexc(\J_k)+\mathcal{J}_\indexh(\J_k)\right]$ is the corresponding power. The prefactor in Eq.~\eqref{eq_J_jk} is defined as
\begin{equation}\label{eq_F}
  F(\J_k)\coloneq\sum_{j=0}^{2\J_k-1} p_j^{\mathrm{ss},k}(j+1)(2\J_k-j),
\end{equation}
where $p_j^{\mathrm{ss},k}$ is the (thermal) population of the $j$th level of the spin-$\J_k$ particle at inverse temperature $\betaeff$. The explicit form of Eq.~\eqref{eq_F} is given in \ref{app_F}. The structure of the energy currents~\eqref{eq_currents_tot} thus accounts for the fact that the different spins in the decomposition~\eqref{eq_J_Sk} (pertaining to different irreducible subspaces) act as individual, i.e., non-interacting, working media of the engine. Thereby, the contribution of each spin $\J_k$ is weighted by its population in the initial state. These populations are conserved since the irreducible subspaces are dynamically invariant. The sign of each individual contribution~\eqref{eq_J_jk} to the heat currents is determined by the difference of the Boltzmann factors of the bath [at temperature $T_i$ and energy $\hbar(\omega_0+q\Omega)$] and of the spin (at temperature $(\kB\betaeff)^{-1}$ and energy $\hbar\omega_0$). Finally, as will become clear below, the prefactor~\eqref{eq_F} contains the collective enhancement of the energy currents.

\par
\begin{figure}
  \centering
  \includegraphics[width=0.95\columnwidth]{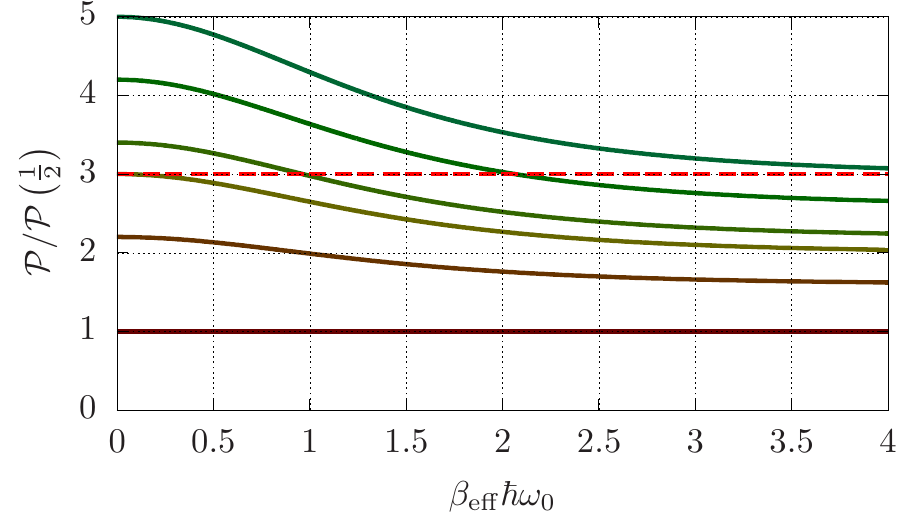}
  \caption{Power~\eqref{eq_P_three_atoms} for $N=3$ atoms and different initial conditions, in units of the power generated by an engine working with a single two-level atom. The initial overlap with the spin-$3/2$ subspace is $\ew{\Pi_1}_{\rho_0}=\{1,0.8,0.6,0.5,0.3,0\}$ (from top to bottom). The horizontal dashed red line is the power generated by three independent two-level atoms. A power above this line indicates a cooperative benefit compared to individual atoms, owing to constructive interference. By contrast, a power below this line shows the detrimental effect of destructive interference.}\label{fig_power_ratio_initial_condition}
\end{figure}
\par

The heat currents~\eqref{eq_J_Sk} demonstrate the crucial role of the initial condition. Consider the case $N=3$ as an illustrative example. As mentioned above, $2\otimes2\otimes2=4\oplus2\oplus2$, such that $\J_1=\frac{3}{2}$, $\J_2=\frac{1}{2}$ and $\J_3=\frac{1}{2}$, respectively. There are thus $n=3$ invariant subspaces and the power~\eqref{eq_P_tot} evaluates to
\begin{equation}\label{eq_P_three_atoms}
  \mathcal{P}=\ew{\Pi_1}_{\rho_0}\mathcal{P}\left(\frac{3}{2}\right)+\left(\ew{\Pi_2}_{\rho_0}+\ew{\Pi_3}_{\rho_0}\right)\mathcal{P}\left(\frac{1}{2}\right).
\end{equation}
As a consequence, for an initial condition having its support only in the two irreducible doublet spaces such that $\ew{\Pi_2}_{\rho_0}+\ew{\Pi_3}_{\rho_0}=1$, the power $\mathcal{P}$ induced by the three two-level atoms equals the power generated by a single two-level atom, $\mathcal{P}=\mathcal{P}(\frac{1}{2})$. Hence, owing to destructive interference, for such an initial condition there is no benefit to, e.g., heat-engine operation, in adding two additional atoms. Figure~\ref{fig_power_ratio_initial_condition} shows the power~\eqref{eq_P_three_atoms} for three atoms and different initial conditions, i.e., different initial weights $\ew{\Pi_k}_{\rho_0}$. As discussed below, maximum cooperativity is achieved under the condition of perfect constructive interference, which is realised by an initial condition in the spin-$N/2$ subspace. Yet, imperfections in the initial preparation of the atoms do not necessarily rule out the possibility of cooperative effects: There may be power enhancement but it will always be smaller than for the ideal case.

\par

Introducing the heat current $\mathcal{J}_i(\frac{1}{2})$ induced by a single two-level atom, the collective heat currents~\eqref{eq_J_jk} can be cast into the simple form
\begin{equation}\label{eq_J_jk_spin}
  \mathcal{J}_i(\J_k)=\frac{F(\J_k)}{F\left(\frac{1}{2}\right)}\mathcal{J}_i\left(\frac{1}{2}\right).
\end{equation}
Since the ratio in Eq.~\eqref{eq_J_jk_spin} does not explicitly depend on the bath index $i$, both heat currents $\mathcal{J}_\indexh(\J_k)$ and $\mathcal{J}_\indexc(\J_k)$ are equally amplified~\cite{niedenzu2015performance} such that also the power originating from the spin-$\J_k$ subspace is equally enhanced with respect to the power generated by a single two-level atom,
\begin{equation}\label{eq_P_jk_spin}
  \mathcal{P}(\J_k)=\frac{F(\J_k)}{F\left(\frac{1}{2}\right)}\mathcal{P}\left(\frac{1}{2}\right).
\end{equation}

\par

The results~\eqref{eq_heat}--\eqref{eq_P_jk_spin} are general in that they do not only apply to heat engines (that convert heat from the hot bath into power) but also to refrigerators (that consume power to cool the cold bath) or any other kind of setups, e.g., heat distributors where both baths are heated up on the expense of the invested power.

\section{Cooperative power enhancement}\label{sec_power}

Example~\eqref{eq_P_three_atoms} demonstrates the crucial role of the atomic initial condition $\rho_0$ in cooperative many-body thermal machines. Perfect constructive interference of the fields scattered by the atoms is achieved under the condition of full atomic cooperativity where all $N$ two-level atoms form the largest-possible spin $N/2$. In order to achieve the maximal benefit from cooperativity it is therefore important to start in a favourable initial condition where only the subspace associated with this spin-$N/2$ is populated, i.e., $\ew{\Pi_K}_{\rho_0}=1$ ($\J_K=\frac{N}{2}$). This subspace is unique and comprises, for example, all $N$ atoms being initially excited or all $N$ atoms being initially in the ground state (see \ref{app_spin_N_half})~\cite{breuerbook}. Note that all the fully-symmetric Dicke states spanning this subspace are eigenstates of the system Hamiltonian and that the preparation of the initial state is a one-time process. Any arbitrary initial condition from this subspace thus guarantees that only the spin-$N/2$ is populated in the state~\eqref{eq_rho_ss}, i.e., that the $N$ two-level atoms act as one collective spin-$N/2$ particle, which is now the relevant working medium of the engine. The specific initial condition of this effective working medium then does not affect the thermodynamic properties of the engine under steady-state operation, which is a periodic limit cycle~\cite{kosloff2013quantum,gelbwaser2013minimal,szczygielski2013markovian,alicki2014quantum,gelbwaser2015thermodynamics,brandner2016periodic}.

\par
\begin{figure}
  \centering
  \includegraphics[width=0.95\columnwidth]{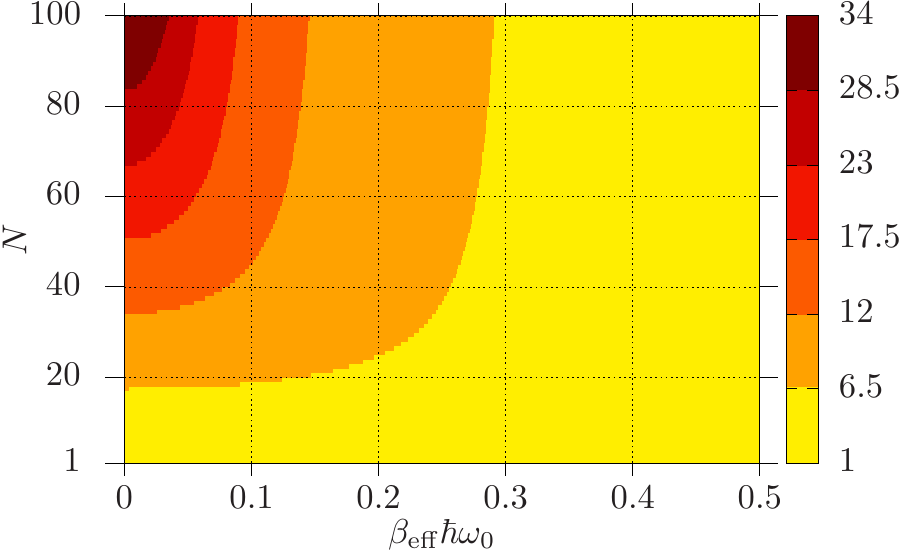}
  \caption{Ratio~\eqref{eq_P_ratio} of the power generated by $N$ collective atoms compared to the power generated by $N$ individual two-level-atom heat machines as a function of the inverse effective temperature~\eqref{eq_betaeff} and the particle number.}\label{fig_power_ratio_map}
\end{figure}
\par

\par
\begin{figure}
  \centering
  \includegraphics[width=0.95\columnwidth]{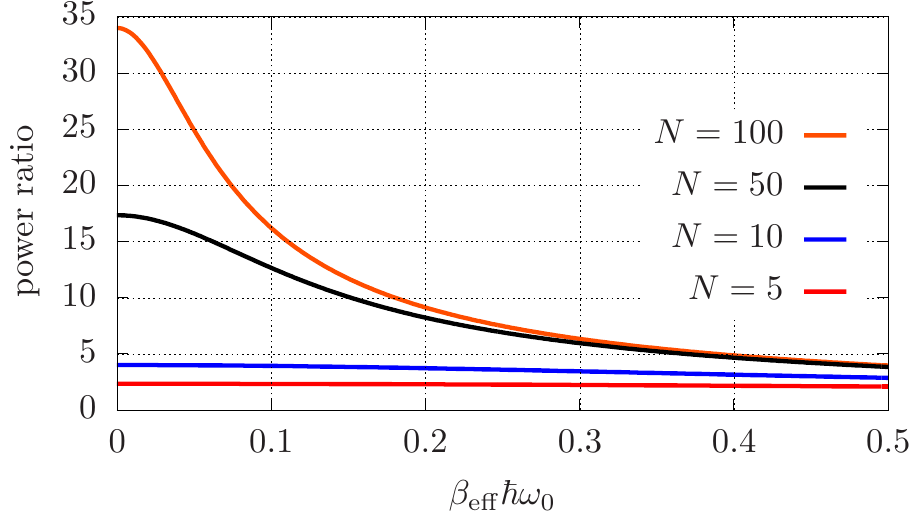}
  \caption{Cuts through Fig.~\ref{fig_power_ratio_map} for different particle numbers $N=\{5,10,50,100\}$ (from bottom to top). The maximum enhancement factor of $(N+2)/3$ is reached for $\betaeff\hbar\omega_0\rightarrow 0$ [Eq.~\eqref{eq_P_ratio_0}].}\label{fig_power_ratio}
\end{figure}
\par

The collective power $\mathcal{P}_\mathrm{coll}\coloneq \mathcal{P}\left(\frac{N}{2}\right)$ and its counterpart $\mathcal{P}_\mathrm{ind}\coloneq N \mathcal{P}\left(\frac{1}{2}\right)$ established by $N$ individual atoms (\cf Fig.~\ref{fig_heat_engine_boost}) then fulfill the relation
\begin{equation}\label{eq_P_ratio}
  \frac{\mathcal{P}_\mathrm{coll}}{\mathcal{P}_\mathrm{ind}}=\frac{1}{N}\frac{F\left(\frac{N}{2}\right)}{F\left(\frac{1}{2}\right)}.
\end{equation}
The ratio~\eqref{eq_P_ratio} is shown in Figs.~\ref{fig_power_ratio_map} and~\ref{fig_power_ratio}. Owing to Eq.~\eqref{eq_J_jk_spin} the same result~\eqref{eq_P_ratio} also holds for the ratio of the collective heat currents $\mathcal{J}_i^\mathrm{coll}\coloneq\mathcal{J}_i(\frac{N}{2})$ to their individual counterpart $\mathcal{J}_i^\mathrm{ind}\coloneq N\mathcal{J}_i(\frac{1}{2})$ ($i\in\{\indexc,\indexh\}$). This implies that there is no collective effect on the engine efficiency since the ratio $\eta=-\mathcal{P}/\mathcal{J}_\indexh$ coincides in the collective and individual cases, $-\mathcal{P}_\mathrm{coll}/\mathcal{J}_\indexh^\mathrm{coll}=-\mathcal{P}_\mathrm{ind}/\mathcal{J}_\indexh^\mathrm{ind}$.

\par

Equation~\eqref{eq_P_ratio} demonstrates the cooperative character of the engine: The collective power $\mathcal{P}_\mathrm{coll}$ is a consequence of quantum coherence between the $N$ two-level atoms. Consequently, destroying this coherence by applying dephasing would cause all the $N$ atoms to interact independently with the baths and the driving field (see~\ref{app_dephasing}). The power output of such an engine with $N$ individual two-level working media equals the power output $\mathcal{P}_\mathrm{ind}$ of $N$ individual engines with a single two-level working medium.

\par

The limiting cases of the ratio~\eqref{eq_P_ratio} for a fixed particle number are
\begin{equation}\label{eq_P_ratio_inf}
  \lim_{\betaeff\hbar\omega_0\rightarrow\infty}\frac{\mathcal{P}_\mathrm{coll}}{\mathcal{P}_\mathrm{ind}}=1
\end{equation}
in the low-temperature regime and
\begin{equation}\label{eq_P_ratio_0}
  \lim_{\betaeff\hbar\omega_0\rightarrow0}\frac{\mathcal{P}_\mathrm{coll}}{\mathcal{P}_\mathrm{ind}}=\frac{N+2}{3}
\end{equation}
in the high-temperature regime. The superradiant scaling behaviour $\mathcal{P}_\mathrm{coll}\sim N\mathcal{P}_\mathrm{ind}=N^2\mathcal{P}(\frac{1}{2})$ is thus established for sufficiently high effective temperatures such that the spin-$N/2$ particle is considerably excited [\cf Eq.~\eqref{eq_rho_ss_k}]. This non-linear scaling with the system size is a direct consequence of cooperativity between the atoms, similar to Dicke superradiance where full cooperativity between the emitters causes an inverted ensemble to radiate in the form of a short burst with peak intensity proportional to $N^2$~\cite{breuerbook}.

\par

For $\betaeff\hbar\omega_0\rightarrow\infty$ the respective energy currents are equal [Eq.~\eqref{eq_P_ratio_inf}] since for such low effective temperatures both the two-level atoms (in the individual case) and the spin-$N/2$ (in the collective case) are mostly in their respective ground state [\cf Eq.~\eqref{eq_rho_ss_k}]. Since, however, these states coincide~\cite{breuerbook} there is no difference between collective or individual atoms and both give rise to the same heat current, $\mathcal{J}_i^\mathrm{coll}=\mathcal{J}_i^\mathrm{ind}$, and thus to the same power, $\mathcal{P}_\mathrm{coll}=\mathcal{P}_\mathrm{ind}$.

\par

By contrast, in the high effective-temperature limit $\betaeff\hbar\omega_0\rightarrow 0$ more and more levels of the spin-$N/2$ become excited. Since the transition probabilities between its individual levels $\ket{j}$ (carrying $j$ excitations; see \ref{app_spin_N_half}) are enhanced by the Clebsch--Gordan coefficients~\cite{breuerbook},
\begin{equation}\label{eq_S_minus_N_half}
  S^{K}_-=\sum_{j=0}^{N-1}\sqrt{(j+1)(N-j)}\ketbra{j}{j+1},
\end{equation}
the coupling of the individual levels to the bath is increased compared to the single two-level case $\sigma_-=\ketbra{g}{e}$. These enhanced transition probabilities play a role once the corresponding levels are populated, which requires a sufficiently high effective temperature in the state~\eqref{eq_rho_ss_k}. This behaviour is also reflected in the amplification function~\eqref{eq_F} where the squared coefficients from Eq.~\eqref{eq_S_minus_N_half} are weighted by the respective thermal populations of the corresponding levels of the spin-$N/2$ particle at inverse temperature $\betaeff$.

\par
\begin{figure}
  \centering
  \includegraphics[width=0.95\columnwidth]{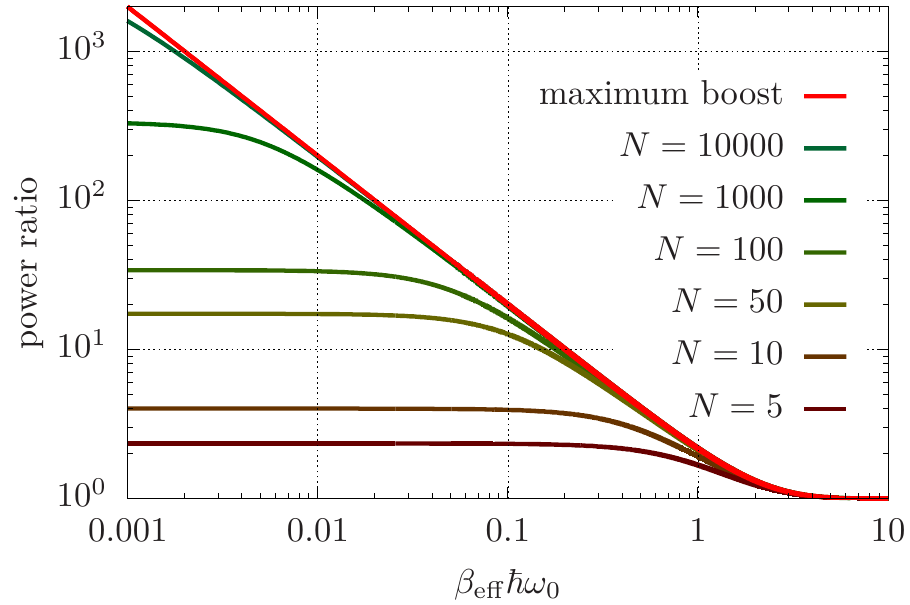}
  \caption{Power boost~\eqref{eq_P_ratio} for different atom numbers and maximum (saturation) power boost~\eqref{eq_P_ratio_coth} achievable for a given $\betaeff\hbar\omega_0$ by increasing the number of atoms. The saturation value~\eqref{eq_P_ratio_coth} cannot be surpassed even by adding more atoms. For example, for $\betaeff\hbar\omega_0=0.2$ the power can only be boosted by a factor of $10$ even if there were thousands of cooperative atoms.}\label{fig_coth}
\end{figure}
\par

Conversely, for a given value of $\betaeff$ (which depends on the modulation~$\omega(t)$ and the bath properties) we find the saturation relation
\begin{equation}\label{eq_P_ratio_coth}
  \lim_{N\rightarrow\infty}\frac{\mathcal{P}_\mathrm{coll}}{\mathcal{P}_\mathrm{ind}}=\coth\left(\frac{\betaeff\hbar\omega_0}{2}\right).
\end{equation}
Again, as expected from Fig.~\ref{fig_power_ratio}, in the low effective-temperature regime $\betaeff\hbar\omega_0\gg 1$ the r.h.s.\ of Eq.~\eqref{eq_P_ratio_coth} tends to unity, such that even for large particle numbers no power enhancement compared to the individual-atom case can be achieved. By contrast, in the high-temperature regime $\betaeff\hbar\omega_0\rightarrow 0$ the r.h.s.\ of Eq.~\eqref{eq_P_ratio_coth} diverges as $2(\betaeff\hbar\omega_0)^{-1}$, such that collective effects significantly enhance the power in this parameter regime (see Fig.~\ref{fig_coth}). The saturation value~\eqref{eq_P_ratio_coth} is then extremely sensitive to the value of $\betaeff\hbar\omega_0$. Hence, there is always a maximum (saturation) atom number above which the power is not amplified further. Consequently, the ideal superradiant scaling behaviour~\eqref{eq_P_ratio_0} for all $N$ is, strictly speaking, only achieved in the strict limit $\betaeff\hbar\omega_0\rightarrow 0$.

\par

Dicke superradiance is commonly related to intense short light pulses emitted by a collection of inverted two-level atoms~\cite{mandelbook,breuerbook,gross1982superradiance}. By contrast, here we rather have ``continuous'' or ``persistent'' superradiance (see also Ref.~\cite{meiser2010steady}). Under steady-state operation photons are continuously exchanged between the baths, the power source and the atoms, which gives rise to collectively-enhanced steady-state energy flows (\cf Fig.~\ref{fig_heat_engine_boost}).

\section{Example: Cooperative energy currents in a heat engine}\label{sec_sin}

As an example, we consider a sinusoidal frequency modulation of the form
\begin{equation}\label{eq_sin}
  \omega(t)=\omega_0+g\sin(\Omega t)
\end{equation}
in the Hamiltonian~\eqref{eq_H_S} under the condition $0\leq g\ll \Omega \leq \omega_0$~\cite{gelbwaser2013minimal,alicki2014quantum}. For this modulation the weights $P(q)$ in the master equation~\eqref{eq_master} evaluate to~\cite{alicki2014quantum}
\begin{subequations}\label{eq_P_sin}
  \begin{align}
    P(0)&\simeq 1-\frac{1}{2}\left(\frac{g}{\Omega}\right)^2\\
    P(\pm1)&\simeq \left(\frac{g}{2\Omega}\right)^2,
  \end{align}
\end{subequations}
such that the higher Floquet sidebands $|q|>1$ do not contribute significantly and can therefore be neglected.

\par

Heat-engine or refrigeration operation (Fig.~\ref{fig_thermal_machines}), respectively, require the photons that are exchanged between the atoms and the hot bath to carry more energy than the photons exchanged between the atoms and the cold bath. This requirement can be realised by spectrally-separated baths, e.g., by imposing~\cite{gelbwaser2013minimal,alicki2014quantum}
\begin{subequations}\label{eq_G_sin}
  \begin{align}
    G_\indexc(\omega)&\approx 0 \quad\text{for}\quad\omega\geq\omega_0\\
    G_\indexh(\omega)&\approx 0 \quad\text{for}\quad\omega\leq\omega_0.
  \end{align}
\end{subequations}

\par
\begin{figure}
  \centering
  \includegraphics[width=\columnwidth]{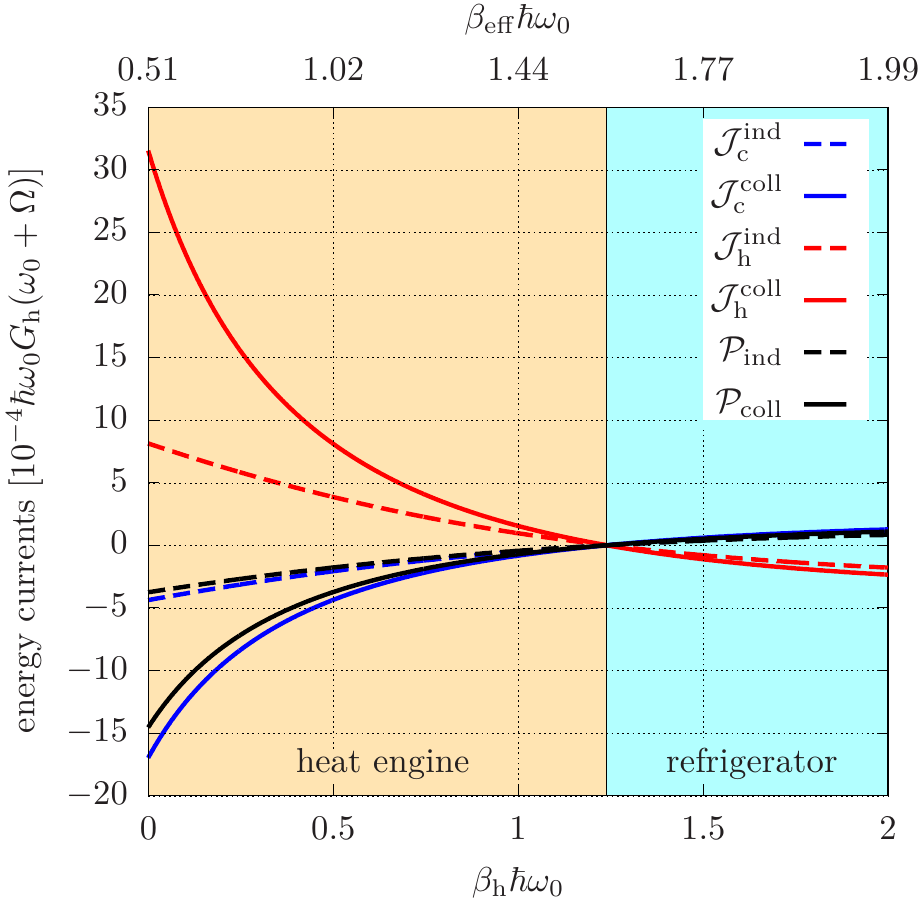}
  \caption{Energy currents generated by $N=100$ individual atoms (dashed) and collective atoms (solid) as a function of $\betah\hbar\omega_0$ under the conditions~\eqref{eq_P_sin} and~\eqref{eq_G_sin}. The maximum power boost of $\mathcal{P}_\mathrm{coll}/\mathcal{P}_\mathrm{ind}\approx 4$ is attained for $\betah\hbar\omega_0\rightarrow 0$, i.e., for a very high temperature of the hot bath (\cf Fig.~\ref{fig_power_ratio} for $\betaeff\hbar\omega_0\approx 0.51$). This maximum boost cannot be increased by adding more atoms since the saturation value~\eqref{eq_P_ratio_coth} also evaluates to $4$ (\cf Fig.~\ref{fig_coth}). Note that adding dephasing would destroy the coherence between the atoms and reduce the collective energy currents $\{J_i^\mathrm{coll},\mathcal{P}_\mathrm{coll}\}$ to their counterparts $\{J_i^\mathrm{ind},\mathcal{P}_\mathrm{ind}\}$ generated by $N$ individual atoms (see~\ref{app_dephasing}). Parameters: $\exp(-\betac\hbar\omega_0)=0.1$ (corresponding to $\betac\hbar\omega_0=2.3$), $\Omega=0.3\omega_0$, $G_\indexc(\omega_0-\Omega)=G_\indexh(\omega_0+\Omega)$ and $g=0.01\Omega$.}\label{fig_currents_Tc_cold}
\end{figure}
\par

\par
\begin{figure}
  \centering
  \includegraphics[width=\columnwidth]{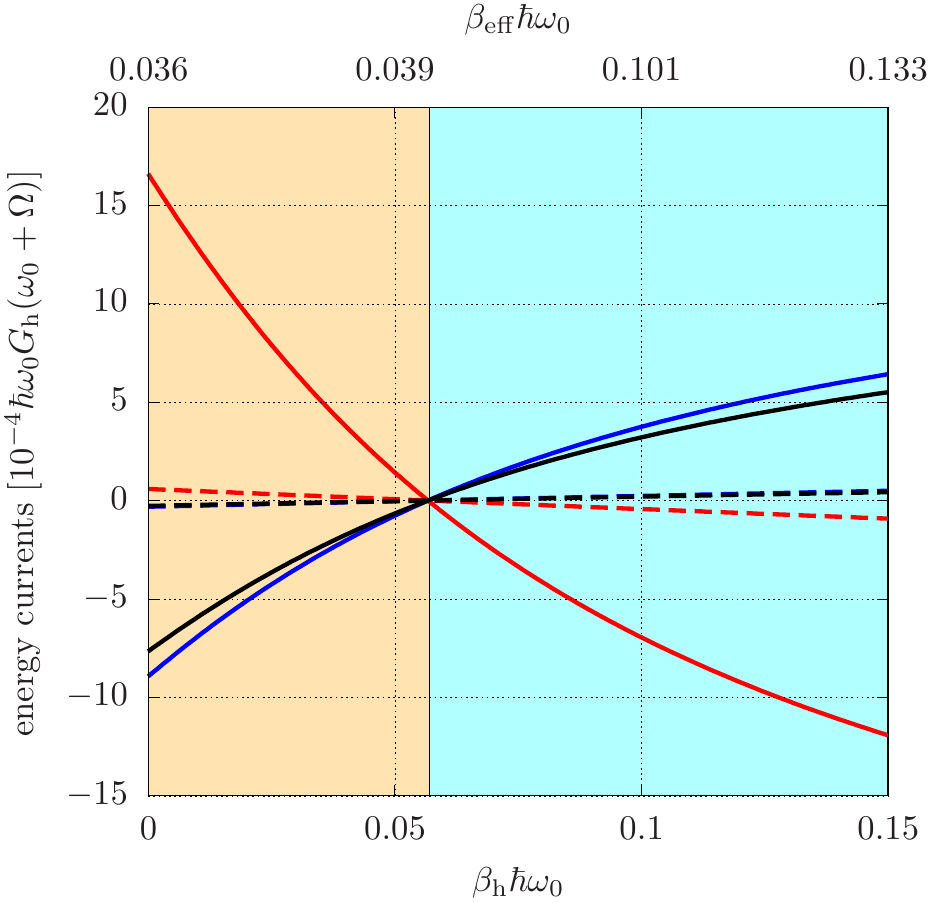}
  \caption{Same as Fig.~\ref{fig_currents_Tc_cold} for a higher temperature of the cold bath, $\exp(-\betac\hbar\omega_0)=0.9$ (corresponding to $\betac\hbar\omega_0=0.11$). Here the minimal value of $\betaeff\hbar\omega_0$ is $0.036$, which results in a power boost of $\mathcal{P}_\mathrm{coll}/\mathcal{P}_\mathrm{ind}\approx 28$ (\cf Fig.~\ref{fig_power_ratio}). Contrary to the situation in Fig.~\ref{fig_currents_Tc_cold} there is a benefit in adding more atoms since here the saturation value~\eqref{eq_P_ratio_coth} evaluates to roughly $56$ (\cf Fig.~\ref{fig_coth}). Owing to the smaller gradient between the two bath temperatures, the power is less than in Fig.~\ref{fig_currents_Tc_cold}. The cooperative power boost, however, is larger. Note the different scaling of the axes compared to Fig.~\ref{fig_currents_Tc_cold}. As in Fig.~\ref{fig_currents_Tc_cold}, adding dephasing would reduce the collective energy currents to their counterparts for individual atoms.}\label{fig_currents_Tc_hot}
\end{figure}
\par

For the sinusoidal modulation~\eqref{eq_sin} and the spectral conditions~\eqref{eq_G_sin}, the thermal machine described by the master equation~\eqref{eq_master} can on-demand act either as a heat engine or a refrigerator~\cite{gelbwaser2013minimal,alicki2014quantum}. Figures~\ref{fig_currents_Tc_cold} and~\ref{fig_currents_Tc_hot} show the energy currents generated by $100$ individual atoms compared to their collective counterparts (the explicit expressions for the currents are given in \ref{app_sin}).

\par

These results imply the following conclusions: The largest possible power boost~\eqref{eq_P_ratio_0}---i.e., the largest-possible collective quantum many-body effect---requires the two bath temperatures to be similar and sufficiently large compared to $\omega_0$, $\betac\hbar\omega_0\ll1$ and $\betah\hbar\omega_0\ll1$ such that $\betaeff\hbar\omega_0\rightarrow 0$. The steady state~\eqref{eq_rho_ss_k} of the spin-$N/2$ particle is then the maximum-entropy state with equally-populated energy levels. The quantitative values of the energy currents may, however, be small despite the superradiant boost (see Fig.~\ref{fig_currents_Tc_hot}).

\par

If, however, the cold-bath temperature is predetermined and the task is to produce large power, irrespective of how large the collective boost is, then it is more favourable to simply increase the hot-bath temperature as much as possible so as to generate the largest possible temperature gradient, as expected for any heat engine~\cite{cengelbook}. Even though the collective effect may then be small, it will still be present (see Fig.~\ref{fig_currents_Tc_cold}).

\par

In general, for predetermined finite bath temperatures it is favourable to strive for a collective behaviour of the atoms by imposing the symmetry that leads to the establishment of invariant subspaces, provided that the initial condition is carefully chosen. The SU(2) symmetry that manifests itself in the establishment of invariant subspaces is thus a thermodynamic resources that can be exploited to increase the output power of heat engines (or the cooling current of refrigerators) compared to an incoherent additive (and hence, in a sense, classical) ensemble of individual heat engines. Superradiance is established by entanglement between the atoms and is thus a genuine quantum effect~\cite{mandelbook}.

\par

In summary, the maximum boost~\eqref{eq_P_ratio_0} is extremely sensitive to the bath temperatures (see Figs.~\ref{fig_power_ratio_map}--\ref{fig_coth} where for small $\betaeff\hbar\omega_0$ and large atom numbers the boost increases very steeply). Consequently, even though the parameters in Fig.~\ref{fig_currents_Tc_hot} allow for very small values of $\betaeff\hbar\omega_0$, the maximum enhancement~\eqref{eq_P_ratio_coth} still saturates to roughly $56$ (\cf Fig.~\ref{fig_coth}), which, depending on the atom number $N$, may be far away from the maximum value $(N+2)/3$ that is attained in the limit $\betaeff\hbar\omega_0\rightarrow0$ [Eq.~\eqref{eq_P_ratio_0}]. In general, the latter limit requires both baths to have comparable temperatures much higher than the bare atomic transition frequency $\omega_0$ such that the absolute value of the power decreases significantly compared to the case of distinct temperatures shown in Fig.~\ref{fig_currents_Tc_cold}.

\par

Hence, there is a severe tradeoff between the realisation of a large quantum boost and the generation of a large power output. Depending on what feature is to be achieved in an experiment, the operation point of the machine must be chosen accordingly.

\section{Collective refrigeration}\label{sec_refrigerator}

The ratio~\eqref{eq_P_ratio} of the collective energy currents to their individual counterparts is universally valid, independent of whether the thermal machine is operated as an engine or a refrigerator (\cf Fig.~\ref{fig_thermal_machines}). Hence, collective effects are not only favourable to engine operation but can also enhance the performance of a refrigerator (i.e., the cold current $\Jc$). This enhancement, however, comes to the price of an equally-scaled larger external power supply.

\par

As seen from Figs.~\ref{fig_currents_Tc_cold} and~\ref{fig_currents_Tc_hot}, the heat flow $\Jc$ increases the colder the hot bath becomes. The largest boost, however, occurs for the smallest possible $\betaeff\hbar\omega_0$ (see Fig.~\ref{fig_power_ratio_map}), which is the regime where the energy currents vanish (see \ref{app_sin}). Just like for the engine case we can conclude that for given bath temperatures there is always a benefit of collective operation (\cf Fig.~\ref{fig_power_ratio}). However, whereas for the engine the boosted power follows from ``free'' enhanced heat currents, the enhanced refrigerator requires the cost of an equally-scaled larger power investment compared to the individual-atoms case.

\section{Realisation considerations}\label{sec_realisation}

The collective behaviour that results in the enhancement of the steady-state energy currents is a consequence of the indistinguishably of the atoms with respect to the two baths. This implies an SU(2) symmetry and the existence of invariant irreducible subspaces. The coupling Hamiltonian~\eqref{eq_H_SB} thus describes the optimal (ideal) situation. Fortunately, its strict symmetry is not necessarily required to obtain a symmetry-preserving Markovian master equation of the form~\eqref{eq_master}.

\par

Consider, e.g., the interaction of $N$ two-level atoms with the surrounding electric field in dipole approximation~\cite{lehmberg1970radiation},
\begin{align}\label{eq_d_E}
  H_\mathrm{SB}&=-\sum_{j=1}^N\mathbf{d}_j\cdot\mathbf{E}\notag\\&=-\sum_{j=1}^N\sum_{\mathbf{k}}f_\mathbf{k}\sigma_x\otimes \left(b_\mathbf{k}e^{i\mathbf{k}\cdot\mathbf{x}_j}+b_\mathbf{k}^\dagger e^{-i\mathbf{k}\cdot\mathbf{x}_j}\right),
\end{align}
where $\mathbf{x}_j$ is the position of the $j$th atom, $f_\mathbf{k}$ the atom--field-mode coupling and $b_\mathbf{k}$ annihilates a photon in mode $\mathbf{k}$ of the electric field. Owing to the position-dependent phases, the operator~\eqref{eq_d_E} cannot be cast in the form~\eqref{eq_H_J_HSB}. Under the Markovian approximation that leads to the master equation~\eqref{eq_master}, however, the atoms effectively only interact with the resonant bath mode (at frequency $\omega_0$ without modulation~\cite{breuerbook} and at frequencies $\omega_0+q\Omega$ with modulation~\cite{szczygielski2014application}, respectively).

\par

In one-dimensional geometries, the atoms can hence effectively appear indistinguishable to the resonant bath mode if they are placed at interatomic distances of integer multiples of the resonance wavelength $\lambda_0=2\pi c/\omega_0$, such that all atoms scatter in phase~\cite{lalumiere2013input,vanloo2013photon,gonzalez2015deterministic}. Since, however, the atoms not only exchange photons with the bath at the bare transition frequency $\omega_0$ but also at the Floquet sidebands [Eq.~\eqref{eq_master}], one must additionally require $2\pi c/(\omega_0+q\Omega)\approx 2\pi c/\omega_0$ for the relevant $q$ (in the example of Sec.~\ref{sec_sin} these are $q=\pm 1$). Hence, under the Markov approximation the dipole Hamiltonian~\eqref{eq_d_E} may effectively be replaced by the fully-symmetric Hamiltonian~\eqref{eq_H_SB}.

\par

The feasibility of superradiance under the interaction~\eqref{eq_d_E} in three dimensions has been experimentally demonstrated for atoms confined within a volume much smaller than the cubed wavelength~\cite{breuerbook,gross1982superradiance}. Then, however, the bath-induced dipole--dipole interaction diverges~\cite{lehmberg1970radiation}. By contrast, it has been theoretically derived~\cite{lalumiere2013input} and experimentally demonstrated~\cite{vanloo2013photon} that in one dimension maximal collective coupling to the bath can be realised while entirely suppressing the bath-induced dipole--dipole interaction by placing the atoms in a chain at distances $d=\lambda_0$ or integer multiples thereof.

\par

The above general requirement of the atoms appearing identical to the bath at all Floquet sidebands can be lifted by imposing a spectral separation of the two baths as introduced in Sec.~\ref{sec_sin} for machine operation. For the modulation~\eqref{eq_sin} and the spectral separation~\eqref{eq_G_sin} the two baths may be realised by two crossed bad cavities (similar to the experiment~\cite{leonard2017supersolid}) or waveguides at different temperatures $\Tc$ and $\Th$ and different resonance frequencies $\omega_0\pm\Omega$. Another possibility may be to consider two commensurable modes of a bad cavity~\cite{keller2018quenches} at different temperatures where the free spectral range of the cavity is $2\Omega$, similar to the two-mode Tavis--Cummings model~\cite{kopylov2015dissipative,moodie2018generalized}. The feasibility of collective strong coupling in multimode cavities has been demonstrated experimentally~\cite{wickenbrock2013collective}. Dicke superradiance has also been experimentally realised for superconducting qubits in microwave cavities~\cite{mlynek2014observation} and for atoms placed along a photonic crystal waveguide~\cite{goban2015superradiance}. Another possible implementation may be in nitrogen-vacancy (NV) centres in diamond nanocrystals, where the recent room-temperature superradiance experiment~\cite{bradac2017room} is close to our assumptions.

\section{Conclusions}\label{sec_conclusions}

We have investigated the impact of cooperative many-body effects on the operation of quantum thermal machines. In suitable geometries and for a carefully-chosen initial condition, a collection of $N$ two-level atoms may behave as a giant collective spin-$N/2$ particle whose energy levels are entangled many-body states. The underlying SU(2) symmetry is manifested in the conservation of the collective spin. This behaviour leads to significantly-enhanced energy currents and to a non-extensive scaling of the power output. Namely, the power generated by a quantum heat engine that involves $N$ collective atoms in its working medium surpasses the power generated by $N$ individual engines that operate with a single-atom working medium. We have mapped the thermal machine with $N$ two-level atoms on a machine with a single, \emph{fictitious}, spin $N/2$-particle. Consequently, for a given $N$, an alternative machine with a single, \emph{physical}, spin-$N/2$ particle would exhibit the same performance as the discussed collective machine. From the point of view of the engine, its working medium is always a spin-$N/2$ in the diagonal state~\eqref{eq_rho_ss}. Hence, although there is coherence in the $N$-atom state, there is no coherence in the physically relevant spin-$N/2$ working medium that otherwise would reduce the power output~\cite{brandner2016periodic,brandner2017universal}.

\par

Whilst we have here considered the case of a modulated transition frequency [Eq.~\eqref{eq_H_S}], our results also apply to different kinds of driving Hamiltonians since our study is based on a general thermodynamic framework~\cite{szczygielski2013markovian,kosloff2013quantum,gelbwaser2013minimal,alicki2014quantum,gelbwaser2015thermodynamics} for periodically-driven open quantum systems. Notably, the Floquet--Bloch master equation for strongly-driven two-level atoms evaluates to a form similar to Eq.~\eqref{eq_master} in the dressed states~\cite{geva1995relaxation,szczygielski2013markovian}, such that similar enhancement features are to be expected in that case too. Our approach is general in that it may be applied to heat engines, refrigerators or any other scenarios like heat distributors that, e.g., occur for the laser-driven cooling of a dephasing bath~\cite{szczygielski2013markovian,gelbwaser2015laser,venkatesh2018cooperative}. We note that although our analysis is based on a general thermodynamic framework~\cite{kosloff2013quantum,gelbwaser2015thermodynamics} that involves a Floquet master equation, we expect the general conclusions to also hold outside the Floquet formalism.

\par

The boosted energy currents are due to ``continuous'' or ``persistent'' superradiance (see also Ref.~\cite{meiser2010steady}). Whereas the power $\mathcal{P}_\mathrm{ind}$ generated by a heat engine that involves $N$ individual two-level atoms as its working medium is, in a sense, classical as it is additive and does not involve quantum-interference effects, the collective power $\mathcal{P}_\mathrm{coll}$ generated under perfect constructive-interference conditions is a genuine quantum many-body effect that involves entanglement between the two-level atoms.

\section*{Acknowledgments}

We would like to thank Arnab Ghosh and Laurin Ostermann for helpful discussions. G.\,K.\ thanks the ISF and DFG for support. W.\,N.\ acknowledges support from an ESQ fellowship of the Austrian Academy of Sciences (\"OAW).

\appendix

\section{Irreducible spin-$N/2$ subspace}\label{app_spin_N_half}

The irreducible spin-$N/2$ subspace is spanned by \mbox{$N+1$} fully-symmetric (under particle exchange) states $\ket{j}$ that carry $j$ excitations. In terms of the $N$-atom states, these basis states read~\cite{breuerbook}
\begin{subequations}
  \begin{align}
    \ket{0}&=\ket{ggg\dots g}\\
    \ket{1}&=S\ket{egg\dots g}\\
    \ket{2}&=S\ket{eeg\dots g}\\
           &\vdots\notag\\
    \ket{N}&=\ket{eee\dots e}.
  \end{align}
\end{subequations}
Here $\ket{g}$ is the single-atom ground state, $\ket{e}$ the excited state and $S$ the symmetrisation operator.

\section{Master equation}\label{app_master}

For $N=1$ the master equation is derived in Sec.~4.1 of Ref.~\cite{alicki2014quantum}. As mentioned in the main text, this derivation also holds for $N>1$ upon replacing the single-atom operators $\sigma_j$ by their cooperative counterparts $J_j$ according to Eqs.~\eqref{eq_J}.

\par

The effective coupling strengths $P(q)$ in Eq.~\eqref{eq_L} are determined by the modulation form $\omega(t)$ as~\cite{alicki2014quantum}
\begin{equation}
  P(q)=\left|\frac{1}{\tau}\int_0^\tau\exp\left(i\int_0^t\left[\omega(s)-\omega_0\right]\dd s\right)e^{-iq\Omega t}\dd t\right|^2,
\end{equation}
where $\tau=2\pi/\Omega$ is the periodicity of the Hamiltonian~\eqref{eq_H_S}, $H_\mathrm{S}(t+\tau)=H_\mathrm{S}(t)$. These coefficients are normalised, $\sum_{q\in\mathbb{Z}}P(q)=1$, and fulfill $P(q)=P(-q)$.

\par

The spectrum $G_i(\omega)$ of the $i$th bath ($i\in\{\indexc,\indexh\}$) at frequency $\omega$ is given by~\cite{breuerbook,alicki2014quantum}
\begin{equation}
  G_i(\omega)=\int_{-\infty}^\infty \ew{B_i(t)B_i(0)}e^{i\omega t}\dd t.
\end{equation}
For bosonic baths the latter evaluates to~\cite{breuerbook,wallsbook}
\begin{equation}
  G_i(\omega)=\gamma_i(\omega)\left[\nbar_i(\omega)+1\right],
\end{equation}
where $\gamma_i(\omega)$ is the spontaneous-emission rate and $\nbar_i(\omega)=[\exp(\beta_i\hbar\omega)-1]^{-1}$ the thermal population of the bath mode at frequency $\omega$. 

\par

The Markovian approximation requires the bath autocorrelation time to be much shorter than the relaxation time of the system~\cite{breuerbook}. Consequently, it is expected to break down for very large $N$~\cite{gross1982superradiance}, for large separations in one-dimensional setups~\cite{lalumiere2013input} or for the spontaneous emission in large atom clouds~\cite{svidzinsky2008dynamical,svidzinsky2008fermi,scully2009super,svidzinsky2010cooperative}. If the assumptions are justified, Markovian master equations can faithfully reproduce superradiance experiments~\cite{wang2007superradiance,vanloo2013photon,mlynek2014observation,araujo2016superradiance,bradac2017room}.

\par

Note that the decay rates from level $\ket{j,m}$ to level $\mapsto\ket{j,m-1}$ ($m=-j,\dots,j$) of the collective spin are~\cite{breuerbook}
  \begin{equation}
    \gamma(m)\propto\bkewabs{j,m-1}{J_-}{j,m}^2=(j-m+1)(j+m).
  \end{equation}
  Hence, the transition rates between the individual levels of the collective spin increase as we approach the central level, which for maximum cooperativity ($j=N/2$) exhibits the superradiant scaling with $N^2$. This implies that for an initially inverted ensemble in contact with a bath at $T=0$ the mean-energy decay rate $\hbar\omega_0\partial\langle J_z\rangle/\partial t$ first rapidly grows and afterwards diminishes~\cite{breuerbook}.

\par

In our treatment we have assumed a perfectly-symmetric Hamiltonian such that the irreducible subspaces are dynamically invariant. Only then does the master equation~\eqref{eq_master} solely involve the collective operators $J_\pm$. If this symmetry is not fulfilled, the master equation will then involve ``cross-decay'' and ``cross-absorption'' terms and be of the form
  \begin{multline}
    \dot\rho=\sum_{i,j=1}^Nc_{ij}(2\sigma_-^i\rho\sigma_+^j-\sigma_+^j\sigma_-^i\rho-\rho\sigma_+^j\sigma_-^i)\\+\sum_{i,j=1}^Nc_{ij}e^{-\betaeff\hbar\omega_0}(2\sigma_+^i\rho\sigma_-^j-\sigma_-^j\sigma_+^i\rho-\rho\sigma_-^j\sigma_+^i),
  \end{multline}
  unless the eigenvalues of the $c_{ij}$ matrix are all zero except for a single eigenvalue which corresponds to collective decay or absorption via the collective-spin operators $J_\mp$ (fulfilled if all the $c_{ij}$ are equal)~\cite{breuerbook}. For small imperfections we can still expect only one eigenvalue of $c_{ij}$ to be dominant; the corresponding decay/absorption channel being $J_\mp$ to a good approximation~\cite{breuerbook}. The steady-state solution~\eqref{eq_rho_ss} is then not unique and depends on the initial condition.

\par

Another kind of imperfections may result in partial symmetry, i.e., some eigenvalues of $c_{ij}$ being zero. Since the determinant of $c_{ij}$ is still zero, the steady-state solution is again initial-condition dependent. In contrast to the ideal case of only one non-zero eigenvalue, now only a sub-ensemble of the $N$ atoms will behave collectively and hence only the initial condition of those atoms appears in the steady-state solution.

\par

By contrast, a broken symmetry, i.e., all eigenvalues of $c_{ij}$ being non-zero, results in an engine run by $N$ independent two-level working media, devoid of collective effects, with a unique steady state of the atoms (similar to the addition of dephasing discussed in~\ref{app_dephasing}). The non-uniqueness of the steady state~\eqref{eq_rho_ss} is thus crucial for the establishment of collective effects---only then are the two-level atoms correlated and not acting independently. As mentioned above, for the ideal case of $N$-atom cooperativity this requires that all the $c_{ij}$ be equal (at least to a good approximation)~\cite{breuerbook}.

\section{Explicit form of $F(\J_k)$}\label{app_F}

Inserting the (thermal) population of the $j$th level of the spin-$\J_k$ particle in state~\eqref{eq_rho_ss_k},
\begin{equation}
  p_j^{\mathrm{ss},k}\coloneq\bkew{k,j}{\rho_\mathrm{ss}^k}{k,j},
\end{equation}
the function~\eqref{eq_F} evaluates to
\begin{equation}
  F(\J_k)=\frac{\sum_{j=0}^{2\J_k-1} e^{-j\betaeff\hbar\omega_0}(j+1)(2\J_k-j)}{\sum_{j=0}^{2\J_k} e^{-j\betaeff\hbar\omega_0}}.
\end{equation}

\section{Dephasing effects}\label{app_dephasing}

Under the influence of local dephasing, the additional symmetry-breaking Liouvillian~\cite{wallsbook}
\begin{equation}
  \mathcal{L}_\mathrm{d}\rho=\sum_{k=1}^N\gamma_\mathrm{d}\mathcal{D}[\sigma_z^k]
\end{equation}
is added to the master equation~\eqref{eq_master}, where $\gamma_\mathrm{d}$ is the dephasing rate (for simplicity assumed to be equal for all atoms). The steady-state solution of the master equation
\begin{equation}
  \dot{\rho}=\sum_{i\in\{\indexc,\indexh\}}\sum_{q\in\mathbb{Z}}\mathcal{L}_{i,q}\rho+\mathcal{L}_\mathrm{d}\rho
\end{equation}
then reads
\begin{equation}
  \rho_\mathrm{ss}^\mathrm{d}=Z^{-1}\exp\left(-\betaeff\frac{\hbar\omega_0}{2}\sum_{k=1}^N\sigma_z^k\right).
\end{equation}
Namely, contrary to Eq.~\eqref{eq_rho_ss}, under the influence of dephasing every two-level atom individually attains a Gibbs-like state at inverse temperature $\betaeff$. For this state the heat currents from the two baths to the atoms evaluate to ($i\in\{\indexc,\indexh\}$)~\cite{alicki2014quantum}
\begin{align}
  \mathcal{J}_i^\mathrm{d}&=\sum_{q\in\mathbb{Z}}\hbar(\omega_0+q\Omega)\Tr\left[\left(\mathcal{L}_{i,q}{\rho}_\mathrm{ss}^\mathrm{d}\right)\frac{1}{2}\sum_{k=1}^N\sigma_z^k\right]\notag\\
                         &=N\mathcal{J}_i\left(\frac{1}{2}\right)\equiv\mathcal{J}_i^\mathrm{ind},
\end{align}
such that the power is
\begin{equation}
  \mathcal{P}^\mathrm{d}=-\left[\mathcal{J}_\indexc^\mathrm{d}+\mathcal{J}_\indexh^\mathrm{d}\right]=N\mathcal{P}\left(\frac{1}{2}\right)\equiv\mathcal{P}_\mathrm{ind}.
\end{equation}
Therefore, under the effect of dephasing the heat engine with $N$ two-level working media gives as much power as $N$ independent heat engines with a single two-level working medium. Namely, the added dephasing destroys the coherence between the individual atoms that gives rise to their collective formation of a big spin.

\section{Energy currents for sinusoidal modulation}\label{app_sin}

For the sinusoidal modulation~\eqref{eq_sin} and the spectral conditions~\eqref{eq_G_sin}, the heat currents~\eqref{eq_J_jk} evaluate to
\begin{subequations}\label{eq_app_heat}
  \begin{align}
    \mathcal{J}_\indexc(\J_k)&=F(\J_k)\hbar(\omega_0-\Omega)\left(\frac{g}{2\Omega}\right)^2G_\indexc(\omega_0-\Omega)\notag\\&\quad\times\left[e^{-\betac\hbar(\omega_0-\Omega)}-e^{-\betaeff\hbar\omega_0}\right]\\
    \mathcal{J}_\indexh(\J_k)&=F(\J_k)\hbar(\omega_0+\Omega)\left(\frac{g}{2\Omega}\right)^2G_\indexh(\omega_0+\Omega)\notag\\&\quad\times\left[e^{-\betah\hbar(\omega_0+\Omega)}-e^{-\betaeff\hbar\omega_0}\right]\\
    \mathcal{P}(\J_k)&=-\left[\mathcal{J}_\indexc(\J_k)+\mathcal{J}_\indexc(\J_k)\right]
  \end{align}
\end{subequations}
and Eq.~\eqref{eq_betaeff} simplifies to
\begin{multline}
  \exp\left(-\betaeff\hbar\omega_0\right)=\\\frac{G_\indexc(\omega_0-\Omega)e^{-\betac\hbar(\omega_0-\Omega)}+G_\indexh(\omega_0+\Omega)e^{-\betah\hbar(\omega_0+\Omega)}}{G_\indexc(\omega_0-\Omega)+G_\indexh(\omega_0+\Omega)}.
\end{multline}
The difference of the Boltzmann factors in Eqs.~\eqref{eq_app_heat} determine the sign of the respective heat currents, i.e., whether the machine acts as an engine ($\mathcal{J}_\indexc<0$, $\mathcal{J}_\indexh>0$ and $\mathcal{P}<0$) or a refrigerator ($\mathcal{J}_\indexc>0$, $\mathcal{J}_\indexh<0$ and $\mathcal{P}>0$). The energy currents~\eqref{eq_app_heat} vanish at the inverse critical temperature
\begin{equation}
  \betah^\mathrm{crit}=\betac\frac{\omega_0-\Omega}{\omega_0+\Omega}
\end{equation}
of the hot bath, which for the parameters in Figs.~\ref{fig_currents_Tc_cold} and~\ref{fig_currents_Tc_hot} evaluates to $\hbar\betah^\mathrm{crit}\omega_0\approx 1.24$ and $\hbar\betah^\mathrm{crit}\omega_0\approx 0.06$, respectively.


\begin{thebibliography}{93}%
\makeatletter
\providecommand \@ifxundefined [1]{%
 \@ifx{#1\undefined}
}%
\providecommand \@ifnum [1]{%
 \ifnum #1\expandafter \@firstoftwo
 \else \expandafter \@secondoftwo
 \fi
}%
\providecommand \@ifx [1]{%
 \ifx #1\expandafter \@firstoftwo
 \else \expandafter \@secondoftwo
 \fi
}%
\providecommand \natexlab [1]{#1}%
\providecommand \enquote  [1]{#1}%
\providecommand \bibnamefont  [1]{#1}%
\providecommand \bibfnamefont [1]{#1}%
\providecommand \citenamefont [1]{#1}%
\providecommand \href@noop [0]{\@secondoftwo}%
\providecommand \href [0]{\begingroup \@sanitize@url \@href}%
\providecommand \@href[1]{\@@startlink{#1}\@@href}%
\providecommand \@@href[1]{\endgroup#1\@@endlink}%
\providecommand \@sanitize@url [0]{\catcode `\\12\catcode `\$12\catcode
  `\&12\catcode `\#12\catcode `\^12\catcode `\_12\catcode `\%12\relax}%
\providecommand \@@startlink[1]{}%
\providecommand \@@endlink[0]{}%
\providecommand \url  [0]{\begingroup\@sanitize@url \@url }%
\providecommand \@url [1]{\endgroup\@href {#1}{\urlprefix }}%
\providecommand \urlprefix  [0]{URL }%
\providecommand \Eprint [0]{\href }%
\providecommand \doibase [0]{http://dx.doi.org/}%
\providecommand \selectlanguage [0]{\@gobble}%
\providecommand \bibinfo  [0]{\@secondoftwo}%
\providecommand \bibfield  [0]{\@secondoftwo}%
\providecommand \translation [1]{[#1]}%
\providecommand \BibitemOpen [0]{}%
\providecommand \bibitemStop [0]{}%
\providecommand \bibitemNoStop [0]{.\EOS\space}%
\providecommand \EOS [0]{\spacefactor3000\relax}%
\providecommand \BibitemShut  [1]{\csname bibitem#1\endcsname}%
\let\auto@bib@innerbib\@empty
\bibitem [{\citenamefont {Kosloff}(2013)}]{kosloff2013quantum}%
  \BibitemOpen
  \bibfield  {author} {\bibinfo {author} {\bibfnamefont {R.}~\bibnamefont
  {Kosloff}},\ }\enquote {\bibinfo {title} {Quantum Thermodynamics: A Dynamical
  Viewpoint},}\ \href {\doibase 10.3390/e15062100} {\bibfield  {journal}
  {\bibinfo  {journal} {Entropy}\ }\textbf {\bibinfo {volume} {15}},\ \bibinfo
  {pages} {2100} (\bibinfo {year} {2013})}\BibitemShut {NoStop}%
\bibitem [{\citenamefont {Gelbwaser-Klimovsky}\ \emph
  {et~al.}(2015{\natexlab{a}})\citenamefont {Gelbwaser-Klimovsky},
  \citenamefont {Niedenzu},\ and\ \citenamefont
  {Kurizki}}]{gelbwaser2015thermodynamics}%
  \BibitemOpen
  \bibfield  {author} {\bibinfo {author} {\bibfnamefont {D.}~\bibnamefont
  {Gelbwaser-Klimovsky}}, \bibinfo {author} {\bibfnamefont {W.}~\bibnamefont
  {Niedenzu}}, \ and\ \bibinfo {author} {\bibfnamefont {G.}~\bibnamefont
  {Kurizki}},\ }\enquote {\bibinfo {title} {Thermodynamics of Quantum Systems
  Under Dynamical Control},}\ \href {\doibase 10.1016/bs.aamop.2015.07.002}
  {\bibfield  {journal} {\bibinfo  {journal} {Adv. At. Mol. Opt. Phys.}\
  }\textbf {\bibinfo {volume} {64}},\ \bibinfo {pages} {329} (\bibinfo {year}
  {2015}{\natexlab{a}})}\BibitemShut {NoStop}%
\bibitem [{\citenamefont {Goold}\ \emph {et~al.}(2016)\citenamefont {Goold},
  \citenamefont {Huber}, \citenamefont {Riera}, \citenamefont {del Rio},\ and\
  \citenamefont {Skrzypczyk}}]{goold2016role}%
  \BibitemOpen
  \bibfield  {author} {\bibinfo {author} {\bibfnamefont {J.}~\bibnamefont
  {Goold}}, \bibinfo {author} {\bibfnamefont {M.}~\bibnamefont {Huber}},
  \bibinfo {author} {\bibfnamefont {A.}~\bibnamefont {Riera}}, \bibinfo
  {author} {\bibfnamefont {L.}~\bibnamefont {del Rio}}, \ and\ \bibinfo
  {author} {\bibfnamefont {P.}~\bibnamefont {Skrzypczyk}},\ }\enquote {\bibinfo
  {title} {The role of quantum information in thermodynamics---a topical
  review},}\ \href {\doibase 10.1088/1751-8113/49/14/143001} {\bibfield
  {journal} {\bibinfo  {journal} {J. Phys. A}\ }\textbf {\bibinfo {volume}
  {49}},\ \bibinfo {pages} {143001} (\bibinfo {year} {2016})}\BibitemShut
  {NoStop}%
\bibitem [{\citenamefont {Vinjanampathy}\ and\ \citenamefont
  {Anders}(2016)}]{vinjanampathy2016quantum}%
  \BibitemOpen
  \bibfield  {author} {\bibinfo {author} {\bibfnamefont {S.}~\bibnamefont
  {Vinjanampathy}}\ and\ \bibinfo {author} {\bibfnamefont {J.}~\bibnamefont
  {Anders}},\ }\enquote {\bibinfo {title} {Quantum thermodynamics},}\ \href
  {\doibase 10.1080/00107514.2016.1201896} {\bibfield  {journal} {\bibinfo
  {journal} {Contemp. Phys.}\ }\textbf {\bibinfo {volume} {57}},\ \bibinfo
  {pages} {1} (\bibinfo {year} {2016})}\BibitemShut {NoStop}%
\bibitem [{\citenamefont {Alicki}\ and\ \citenamefont
  {Kosloff}(2018)}]{alicki2018introduction}%
  \BibitemOpen
  \bibfield  {author} {\bibinfo {author} {\bibfnamefont {R.}~\bibnamefont
  {Alicki}}\ and\ \bibinfo {author} {\bibfnamefont {R.}~\bibnamefont
  {Kosloff}},\ }\enquote {\bibinfo {title} {Introduction to Quantum
  Thermodynamics: History and Prospects},}\ \href
  {https://arxiv.org/abs/1801.08314} {\bibfield  {journal} {\bibinfo  {journal}
  {arXiv preprint arXiv:1801.08314}\ } (\bibinfo {year} {2018})}\BibitemShut
  {NoStop}%
\bibitem [{\citenamefont {\c{C}engel}\ and\ \citenamefont
  {Boles}(2015)}]{cengelbook}%
  \BibitemOpen
  \bibfield  {author} {\bibinfo {author} {\bibfnamefont {Y.~A.}\ \bibnamefont
  {\c{C}engel}}\ and\ \bibinfo {author} {\bibfnamefont {M.~A.}\ \bibnamefont
  {Boles}},\ }\href@noop {} {\emph {\bibinfo {title} {Thermodynamics: An
  Engineering Approach}}},\ \bibinfo {edition} {eighth}\ ed.\ (\bibinfo
  {publisher} {McGraw-Hill Education},\ \bibinfo {address} {New York},\
  \bibinfo {year} {2015})\BibitemShut {NoStop}%
\bibitem [{\citenamefont {Scully}\ \emph {et~al.}(2003)\citenamefont {Scully},
  \citenamefont {Zubairy}, \citenamefont {Agarwal},\ and\ \citenamefont
  {Walther}}]{scully2003extracting}%
  \BibitemOpen
  \bibfield  {author} {\bibinfo {author} {\bibfnamefont {M.~O.}\ \bibnamefont
  {Scully}}, \bibinfo {author} {\bibfnamefont {M.~S.}\ \bibnamefont {Zubairy}},
  \bibinfo {author} {\bibfnamefont {G.~S.}\ \bibnamefont {Agarwal}}, \ and\
  \bibinfo {author} {\bibfnamefont {H.}~\bibnamefont {Walther}},\ }\enquote
  {\bibinfo {title} {Extracting Work from a Single Heat Bath via Vanishing
  Quantum Coherence},}\ \href {\doibase 10.1126/science.1078955} {\bibfield
  {journal} {\bibinfo  {journal} {Science}\ }\textbf {\bibinfo {volume}
  {299}},\ \bibinfo {pages} {862} (\bibinfo {year} {2003})}\BibitemShut
  {NoStop}%
\bibitem [{\citenamefont {Scully}\ \emph {et~al.}(2011)\citenamefont {Scully},
  \citenamefont {Chapin}, \citenamefont {Dorfman}, \citenamefont {Kim},\ and\
  \citenamefont {Svidzinsky}}]{scully2011quantum}%
  \BibitemOpen
  \bibfield  {author} {\bibinfo {author} {\bibfnamefont {M.~O.}\ \bibnamefont
  {Scully}}, \bibinfo {author} {\bibfnamefont {K.~R.}\ \bibnamefont {Chapin}},
  \bibinfo {author} {\bibfnamefont {K.~E.}\ \bibnamefont {Dorfman}}, \bibinfo
  {author} {\bibfnamefont {M.~B.}\ \bibnamefont {Kim}}, \ and\ \bibinfo
  {author} {\bibfnamefont {A.}~\bibnamefont {Svidzinsky}},\ }\enquote {\bibinfo
  {title} {Quantum heat engine power can be increased by noise-induced
  coherence},}\ \href {\doibase 10.1073/pnas.1110234108} {\bibfield  {journal}
  {\bibinfo  {journal} {Proc. Natl. Acad. Sci. USA}\ }\textbf {\bibinfo
  {volume} {108}},\ \bibinfo {pages} {15097} (\bibinfo {year}
  {2011})}\BibitemShut {NoStop}%
\bibitem [{\citenamefont {Brandner}\ \emph {et~al.}(2015)\citenamefont
  {Brandner}, \citenamefont {Bauer}, \citenamefont {Schmid},\ and\
  \citenamefont {Seifert}}]{brandner2015coherence}%
  \BibitemOpen
  \bibfield  {author} {\bibinfo {author} {\bibfnamefont {K.}~\bibnamefont
  {Brandner}}, \bibinfo {author} {\bibfnamefont {M.}~\bibnamefont {Bauer}},
  \bibinfo {author} {\bibfnamefont {M.~T.}\ \bibnamefont {Schmid}}, \ and\
  \bibinfo {author} {\bibfnamefont {U.}~\bibnamefont {Seifert}},\ }\enquote
  {\bibinfo {title} {Coherence-enhanced efficiency of feedback-driven quantum
  engines},}\ \href {\doibase 10.1088/1367-2630/17/6/065006} {\bibfield
  {journal} {\bibinfo  {journal} {New J. Phys.}\ }\textbf {\bibinfo {volume}
  {17}},\ \bibinfo {pages} {065006} (\bibinfo {year} {2015})}\BibitemShut
  {NoStop}%
\bibitem [{\citenamefont {Uzdin}\ \emph {et~al.}(2015)\citenamefont {Uzdin},
  \citenamefont {Levy},\ and\ \citenamefont {Kosloff}}]{uzdin2015equivalence}%
  \BibitemOpen
  \bibfield  {author} {\bibinfo {author} {\bibfnamefont {R.}~\bibnamefont
  {Uzdin}}, \bibinfo {author} {\bibfnamefont {A.}~\bibnamefont {Levy}}, \ and\
  \bibinfo {author} {\bibfnamefont {R.}~\bibnamefont {Kosloff}},\ }\enquote
  {\bibinfo {title} {Equivalence of Quantum Heat Machines, and
  Quantum-Thermodynamic Signatures},}\ \href {\doibase
  10.1103/PhysRevX.5.031044} {\bibfield  {journal} {\bibinfo  {journal} {Phys.
  Rev. X}\ }\textbf {\bibinfo {volume} {5}},\ \bibinfo {pages} {031044}
  (\bibinfo {year} {2015})}\BibitemShut {NoStop}%
\bibitem [{\citenamefont {Niedenzu}\ \emph {et~al.}(2015)\citenamefont
  {Niedenzu}, \citenamefont {Gelbwaser-Klimovsky},\ and\ \citenamefont
  {Kurizki}}]{niedenzu2015performance}%
  \BibitemOpen
  \bibfield  {author} {\bibinfo {author} {\bibfnamefont {W.}~\bibnamefont
  {Niedenzu}}, \bibinfo {author} {\bibfnamefont {D.}~\bibnamefont
  {Gelbwaser-Klimovsky}}, \ and\ \bibinfo {author} {\bibfnamefont
  {G.}~\bibnamefont {Kurizki}},\ }\enquote {\bibinfo {title} {Performance
  limits of multilevel and multipartite quantum heat machines},}\ \href
  {\doibase 10.1103/PhysRevE.92.042123} {\bibfield  {journal} {\bibinfo
  {journal} {Phys. Rev. E}\ }\textbf {\bibinfo {volume} {92}},\ \bibinfo
  {pages} {042123} (\bibinfo {year} {2015})}\BibitemShut {NoStop}%
\bibitem [{\citenamefont {Uzdin}(2016)}]{uzdin2016coherence}%
  \BibitemOpen
  \bibfield  {author} {\bibinfo {author} {\bibfnamefont {R.}~\bibnamefont
  {Uzdin}},\ }\enquote {\bibinfo {title} {Coherence-Induced Reversibility and
  Collective Operation of Quantum Heat Machines via Coherence Recycling},}\
  \href {\doibase 10.1103/PhysRevApplied.6.024004} {\bibfield  {journal}
  {\bibinfo  {journal} {Phys. Rev. Applied}\ }\textbf {\bibinfo {volume} {6}},\
  \bibinfo {pages} {024004} (\bibinfo {year} {2016})}\BibitemShut {NoStop}%
\bibitem [{\citenamefont {Friedenberger}\ and\ \citenamefont
  {Lutz}(2017)}]{friedenberger2017quantum}%
  \BibitemOpen
  \bibfield  {author} {\bibinfo {author} {\bibfnamefont {A.}~\bibnamefont
  {Friedenberger}}\ and\ \bibinfo {author} {\bibfnamefont {E.}~\bibnamefont
  {Lutz}},\ }\enquote {\bibinfo {title} {When is a quantum heat engine
  quantum?}}\ \href {\doibase 10.1209/0295-5075/120/10002} {\bibfield
  {journal} {\bibinfo  {journal} {EPL (Europhys. Lett.)}\ }\textbf {\bibinfo
  {volume} {120}},\ \bibinfo {pages} {10002} (\bibinfo {year}
  {2017})}\BibitemShut {NoStop}%
\bibitem [{\citenamefont {Dillenschneider}\ and\ \citenamefont
  {Lutz}(2009)}]{dillenschneider2009energetics}%
  \BibitemOpen
  \bibfield  {author} {\bibinfo {author} {\bibfnamefont {R.}~\bibnamefont
  {Dillenschneider}}\ and\ \bibinfo {author} {\bibfnamefont {E.}~\bibnamefont
  {Lutz}},\ }\enquote {\bibinfo {title} {Energetics of quantum correlations},}\
  \href {\doibase 10.1209/0295-5075/88/50003} {\bibfield  {journal} {\bibinfo
  {journal} {EPL (Europhys. Lett.)}\ }\textbf {\bibinfo {volume} {88}},\
  \bibinfo {pages} {50003} (\bibinfo {year} {2009})}\BibitemShut {NoStop}%
\bibitem [{\citenamefont {Huang}\ \emph {et~al.}(2012)\citenamefont {Huang},
  \citenamefont {Wang},\ and\ \citenamefont {Yi}}]{huang2012effects}%
  \BibitemOpen
  \bibfield  {author} {\bibinfo {author} {\bibfnamefont {X.~L.}\ \bibnamefont
  {Huang}}, \bibinfo {author} {\bibfnamefont {T.}~\bibnamefont {Wang}}, \ and\
  \bibinfo {author} {\bibfnamefont {X.~X.}\ \bibnamefont {Yi}},\ }\enquote
  {\bibinfo {title} {Effects of reservoir squeezing on quantum systems and work
  extraction},}\ \href {\doibase 10.1103/PhysRevE.86.051105} {\bibfield
  {journal} {\bibinfo  {journal} {Phys. Rev. E}\ }\textbf {\bibinfo {volume}
  {86}},\ \bibinfo {pages} {051105} (\bibinfo {year} {2012})}\BibitemShut
  {NoStop}%
\bibitem [{\citenamefont {Abah}\ and\ \citenamefont
  {Lutz}(2014)}]{abah2014efficiency}%
  \BibitemOpen
  \bibfield  {author} {\bibinfo {author} {\bibfnamefont {O.}~\bibnamefont
  {Abah}}\ and\ \bibinfo {author} {\bibfnamefont {E.}~\bibnamefont {Lutz}},\
  }\enquote {\bibinfo {title} {Efficiency of heat engines coupled to
  nonequilibrium reservoirs},}\ \href {\doibase 10.1209/0295-5075/106/20001}
  {\bibfield  {journal} {\bibinfo  {journal} {EPL (Europhys. Lett.)}\ }\textbf
  {\bibinfo {volume} {106}},\ \bibinfo {pages} {20001} (\bibinfo {year}
  {2014})}\BibitemShut {NoStop}%
\bibitem [{\citenamefont {Ro\ss{}nagel}\ \emph {et~al.}(2014)\citenamefont
  {Ro\ss{}nagel}, \citenamefont {Abah}, \citenamefont {Schmidt-Kaler},
  \citenamefont {Singer},\ and\ \citenamefont {Lutz}}]{rossnagel2014nanoscale}%
  \BibitemOpen
  \bibfield  {author} {\bibinfo {author} {\bibfnamefont {J.}~\bibnamefont
  {Ro\ss{}nagel}}, \bibinfo {author} {\bibfnamefont {O.}~\bibnamefont {Abah}},
  \bibinfo {author} {\bibfnamefont {F.}~\bibnamefont {Schmidt-Kaler}}, \bibinfo
  {author} {\bibfnamefont {K.}~\bibnamefont {Singer}}, \ and\ \bibinfo {author}
  {\bibfnamefont {E.}~\bibnamefont {Lutz}},\ }\enquote {\bibinfo {title}
  {Nanoscale Heat Engine Beyond the Carnot Limit},}\ \href {\doibase
  10.1103/PhysRevLett.112.030602} {\bibfield  {journal} {\bibinfo  {journal}
  {Phys. Rev. Lett.}\ }\textbf {\bibinfo {volume} {112}},\ \bibinfo {pages}
  {030602} (\bibinfo {year} {2014})}\BibitemShut {NoStop}%
\bibitem [{\citenamefont {Hardal}\ and\ \citenamefont
  {M{\"u}stecapl{\i}o{\u{g}}lu}(2015)}]{hardal2015superradiant}%
  \BibitemOpen
  \bibfield  {author} {\bibinfo {author} {\bibfnamefont {A.~{\"U}.~C.}\
  \bibnamefont {Hardal}}\ and\ \bibinfo {author} {\bibfnamefont {{\"O}.~E.}\
  \bibnamefont {M{\"u}stecapl{\i}o{\u{g}}lu}},\ }\enquote {\bibinfo {title}
  {Superradiant Quantum Heat Engine},}\ \href {\doibase 10.1038/srep12953}
  {\bibfield  {journal} {\bibinfo  {journal} {Sci. Rep.}\ }\textbf {\bibinfo
  {volume} {5}},\ \bibinfo {pages} {12953} (\bibinfo {year}
  {2015})}\BibitemShut {NoStop}%
\bibitem [{\citenamefont {Da{\u{g}}}\ \emph {et~al.}(2016)\citenamefont
  {Da{\u{g}}}, \citenamefont {Niedenzu}, \citenamefont
  {M{\"u}stecapl{\i}o{\u{g}}lu},\ and\ \citenamefont
  {Kurizki}}]{dag2016multiatom}%
  \BibitemOpen
  \bibfield  {author} {\bibinfo {author} {\bibfnamefont {C.~B.}\ \bibnamefont
  {Da{\u{g}}}}, \bibinfo {author} {\bibfnamefont {W.}~\bibnamefont {Niedenzu}},
  \bibinfo {author} {\bibfnamefont {{\"O}.~E.}\ \bibnamefont
  {M{\"u}stecapl{\i}o{\u{g}}lu}}, \ and\ \bibinfo {author} {\bibfnamefont
  {G.}~\bibnamefont {Kurizki}},\ }\enquote {\bibinfo {title} {Multiatom Quantum
  Coherences in Micromasers as Fuel for Thermal and Nonthermal Machines},}\
  \href {\doibase 10.3390/e18070244} {\bibfield  {journal} {\bibinfo  {journal}
  {Entropy}\ }\textbf {\bibinfo {volume} {18}},\ \bibinfo {pages} {244}
  (\bibinfo {year} {2016})}\BibitemShut {NoStop}%
\bibitem [{\citenamefont {Manzano}\ \emph {et~al.}(2016)\citenamefont
  {Manzano}, \citenamefont {Galve}, \citenamefont {Zambrini},\ and\
  \citenamefont {Parrondo}}]{manzano2016entropy}%
  \BibitemOpen
  \bibfield  {author} {\bibinfo {author} {\bibfnamefont {G.}~\bibnamefont
  {Manzano}}, \bibinfo {author} {\bibfnamefont {F.}~\bibnamefont {Galve}},
  \bibinfo {author} {\bibfnamefont {R.}~\bibnamefont {Zambrini}}, \ and\
  \bibinfo {author} {\bibfnamefont {J.~M.~R.}\ \bibnamefont {Parrondo}},\
  }\enquote {\bibinfo {title} {Entropy production and thermodynamic power of
  the squeezed thermal reservoir},}\ \href {\doibase
  10.1103/PhysRevE.93.052120} {\bibfield  {journal} {\bibinfo  {journal} {Phys.
  Rev. E}\ }\textbf {\bibinfo {volume} {93}},\ \bibinfo {pages} {052120}
  (\bibinfo {year} {2016})}\BibitemShut {NoStop}%
\bibitem [{\citenamefont {T\"urkpen\ifmmode~\mbox{\c{c}}\else \c{c}\fi{}e}\
  and\ \citenamefont {M\"ustecapl\ifmmode \imath \else \i
  \fi{}o\ifmmode~\breve{g}\else \u{g}\fi{}lu}(2016)}]{turkpence2016quantum}%
  \BibitemOpen
  \bibfield  {author} {\bibinfo {author} {\bibfnamefont {D.}~\bibnamefont
  {T\"urkpen\ifmmode~\mbox{\c{c}}\else \c{c}\fi{}e}}\ and\ \bibinfo {author}
  {\bibfnamefont {{\"O}.~E.}\ \bibnamefont {M\"ustecapl\ifmmode \imath \else \i
  \fi{}o\ifmmode~\breve{g}\else \u{g}\fi{}lu}},\ }\enquote {\bibinfo {title}
  {Quantum fuel with multilevel atomic coherence for ultrahigh specific work in
  a photonic Carnot engine},}\ \href {\doibase 10.1103/PhysRevE.93.012145}
  {\bibfield  {journal} {\bibinfo  {journal} {Phys. Rev. E}\ }\textbf {\bibinfo
  {volume} {93}},\ \bibinfo {pages} {012145} (\bibinfo {year}
  {2016})}\BibitemShut {NoStop}%
\bibitem [{\citenamefont {Agarwalla}\ \emph {et~al.}(2017)\citenamefont
  {Agarwalla}, \citenamefont {Jiang},\ and\ \citenamefont
  {Segal}}]{agarwalla2017quantum}%
  \BibitemOpen
  \bibfield  {author} {\bibinfo {author} {\bibfnamefont {B.~K.}\ \bibnamefont
  {Agarwalla}}, \bibinfo {author} {\bibfnamefont {J.-H.}\ \bibnamefont
  {Jiang}}, \ and\ \bibinfo {author} {\bibfnamefont {D.}~\bibnamefont
  {Segal}},\ }\enquote {\bibinfo {title} {Quantum efficiency bound for
  continuous heat engines coupled to noncanonical reservoirs},}\ \href
  {\doibase 10.1103/PhysRevB.96.104304} {\bibfield  {journal} {\bibinfo
  {journal} {Phys. Rev. B}\ }\textbf {\bibinfo {volume} {96}},\ \bibinfo
  {pages} {104304} (\bibinfo {year} {2017})}\BibitemShut {NoStop}%
\bibitem [{\citenamefont {Da\u{g}}\ \emph {et~al.}(2018)\citenamefont
  {Da\u{g}}, \citenamefont {Niedenzu}, \citenamefont {Ozaydin}, \citenamefont
  {M\"ustecapl{\i}o\u{g}lu},\ and\ \citenamefont
  {Kurizki}}]{dag2018temperature}%
  \BibitemOpen
  \bibfield  {author} {\bibinfo {author} {\bibfnamefont {C.~B.}\ \bibnamefont
  {Da\u{g}}}, \bibinfo {author} {\bibfnamefont {W.}~\bibnamefont {Niedenzu}},
  \bibinfo {author} {\bibfnamefont {F.}~\bibnamefont {Ozaydin}}, \bibinfo
  {author} {\bibfnamefont {{\"O}.~E.}\ \bibnamefont {M\"ustecapl{\i}o\u{g}lu}},
  \ and\ \bibinfo {author} {\bibfnamefont {G.}~\bibnamefont {Kurizki}},\
  }\enquote {\bibinfo {title} {Temperature control in cavities by combustion of
  two-atom entanglement},}\ \href {https://arxiv.org/abs/1801.04529} {\bibfield
   {journal} {\bibinfo  {journal} {arXiv preprint arXiv:1801.04529}\ }
  (\bibinfo {year} {2018})}\BibitemShut {NoStop}%
\bibitem [{\citenamefont {Niedenzu}\ \emph {et~al.}(2018)\citenamefont
  {Niedenzu}, \citenamefont {Mukherjee}, \citenamefont {Ghosh}, \citenamefont
  {Kofman},\ and\ \citenamefont {Kurizki}}]{niedenzu2018quantum}%
  \BibitemOpen
  \bibfield  {author} {\bibinfo {author} {\bibfnamefont {W.}~\bibnamefont
  {Niedenzu}}, \bibinfo {author} {\bibfnamefont {V.}~\bibnamefont {Mukherjee}},
  \bibinfo {author} {\bibfnamefont {A.}~\bibnamefont {Ghosh}}, \bibinfo
  {author} {\bibfnamefont {A.~G.}\ \bibnamefont {Kofman}}, \ and\ \bibinfo
  {author} {\bibfnamefont {G.}~\bibnamefont {Kurizki}},\ }\enquote {\bibinfo
  {title} {Quantum engine efficiency bound beyond the second law of
  thermodynamics},}\ \href {\doibase 10.1038/s41467-017-01991-6} {\bibfield
  {journal} {\bibinfo  {journal} {Nat. Commun.}\ }\textbf {\bibinfo {volume}
  {9}},\ \bibinfo {pages} {165} (\bibinfo {year} {2018})}\BibitemShut {NoStop}%
\bibitem [{\citenamefont {Ro{\ss}nagel}\ \emph {et~al.}(2016)\citenamefont
  {Ro{\ss}nagel}, \citenamefont {Dawkins}, \citenamefont {Tolazzi},
  \citenamefont {Abah}, \citenamefont {Lutz}, \citenamefont {Schmidt-Kaler},\
  and\ \citenamefont {Singer}}]{rossnagel2016single}%
  \BibitemOpen
  \bibfield  {author} {\bibinfo {author} {\bibfnamefont {J.}~\bibnamefont
  {Ro{\ss}nagel}}, \bibinfo {author} {\bibfnamefont {S.~T.}\ \bibnamefont
  {Dawkins}}, \bibinfo {author} {\bibfnamefont {K.~N.}\ \bibnamefont
  {Tolazzi}}, \bibinfo {author} {\bibfnamefont {O.}~\bibnamefont {Abah}},
  \bibinfo {author} {\bibfnamefont {E.}~\bibnamefont {Lutz}}, \bibinfo {author}
  {\bibfnamefont {F.}~\bibnamefont {Schmidt-Kaler}}, \ and\ \bibinfo {author}
  {\bibfnamefont {K.}~\bibnamefont {Singer}},\ }\enquote {\bibinfo {title} {A
  single-atom heat engine},}\ \href {\doibase 10.1126/science.aad6320}
  {\bibfield  {journal} {\bibinfo  {journal} {Science}\ }\textbf {\bibinfo
  {volume} {352}},\ \bibinfo {pages} {325} (\bibinfo {year}
  {2016})}\BibitemShut {NoStop}%
\bibitem [{\citenamefont {Klatzow}\ \emph {et~al.}(2017)\citenamefont
  {Klatzow}, \citenamefont {Becker}, \citenamefont {Ledingham}, \citenamefont
  {Weinzetl}, \citenamefont {Kaczmarek}, \citenamefont {Saunders},
  \citenamefont {Nunn}, \citenamefont {Walmsley}, \citenamefont {Uzdin},\ and\
  \citenamefont {Poem}}]{klatzow2017experimental}%
  \BibitemOpen
  \bibfield  {author} {\bibinfo {author} {\bibfnamefont {J.}~\bibnamefont
  {Klatzow}}, \bibinfo {author} {\bibfnamefont {J.~N.}\ \bibnamefont {Becker}},
  \bibinfo {author} {\bibfnamefont {P.~M.}\ \bibnamefont {Ledingham}}, \bibinfo
  {author} {\bibfnamefont {C.}~\bibnamefont {Weinzetl}}, \bibinfo {author}
  {\bibfnamefont {K.~T.}\ \bibnamefont {Kaczmarek}}, \bibinfo {author}
  {\bibfnamefont {D.~J.}\ \bibnamefont {Saunders}}, \bibinfo {author}
  {\bibfnamefont {J.}~\bibnamefont {Nunn}}, \bibinfo {author} {\bibfnamefont
  {I.~A.}\ \bibnamefont {Walmsley}}, \bibinfo {author} {\bibfnamefont
  {R.}~\bibnamefont {Uzdin}}, \ and\ \bibinfo {author} {\bibfnamefont
  {E.}~\bibnamefont {Poem}},\ }\enquote {\bibinfo {title} {Experimental
  demonstration of quantum effects in the operation of microscopic heat
  engines},}\ \href {https://arxiv.org/abs/1710.08716} {\bibfield  {journal}
  {\bibinfo  {journal} {arXiv preprint arXiv:1710.08716}\ } (\bibinfo {year}
  {2017})}\BibitemShut {NoStop}%
\bibitem [{\citenamefont {Klaers}\ \emph {et~al.}(2017)\citenamefont {Klaers},
  \citenamefont {Faelt}, \citenamefont {Imamoglu},\ and\ \citenamefont
  {Togan}}]{klaers2017squeezed}%
  \BibitemOpen
  \bibfield  {author} {\bibinfo {author} {\bibfnamefont {J.}~\bibnamefont
  {Klaers}}, \bibinfo {author} {\bibfnamefont {S.}~\bibnamefont {Faelt}},
  \bibinfo {author} {\bibfnamefont {A.}~\bibnamefont {Imamoglu}}, \ and\
  \bibinfo {author} {\bibfnamefont {E.}~\bibnamefont {Togan}},\ }\enquote
  {\bibinfo {title} {Squeezed Thermal Reservoirs as a Resource for a
  Nanomechanical Engine beyond the Carnot Limit},}\ \href {\doibase
  10.1103/PhysRevX.7.031044} {\bibfield  {journal} {\bibinfo  {journal} {Phys.
  Rev. X}\ }\textbf {\bibinfo {volume} {7}},\ \bibinfo {pages} {031044}
  (\bibinfo {year} {2017})}\BibitemShut {NoStop}%
\bibitem [{\citenamefont {Jiang}(2014)}]{jiang2014enhancing}%
  \BibitemOpen
  \bibfield  {author} {\bibinfo {author} {\bibfnamefont {J.-H.}\ \bibnamefont
  {Jiang}},\ }\enquote {\bibinfo {title} {Enhancing efficiency and power of
  quantum-dots resonant tunneling thermoelectrics in three-terminal geometry by
  cooperative effects},}\ \href {\doibase 10.1063/1.4901120} {\bibfield
  {journal} {\bibinfo  {journal} {J. Appl. Phys.}\ }\textbf {\bibinfo {volume}
  {116}},\ \bibinfo {pages} {194303} (\bibinfo {year} {2014})}\BibitemShut
  {NoStop}%
\bibitem [{\citenamefont {Binder}\ \emph {et~al.}(2015)\citenamefont {Binder},
  \citenamefont {Vinjanampathy}, \citenamefont {Modi},\ and\ \citenamefont
  {Goold}}]{binder2015quantacell}%
  \BibitemOpen
  \bibfield  {author} {\bibinfo {author} {\bibfnamefont {F.~C.}\ \bibnamefont
  {Binder}}, \bibinfo {author} {\bibfnamefont {S.}~\bibnamefont
  {Vinjanampathy}}, \bibinfo {author} {\bibfnamefont {K.}~\bibnamefont {Modi}},
  \ and\ \bibinfo {author} {\bibfnamefont {J.}~\bibnamefont {Goold}},\
  }\enquote {\bibinfo {title} {Quantacell: powerful charging of quantum
  batteries},}\ \href {\doibase 10.1088/1367-2630/17/7/075015} {\bibfield
  {journal} {\bibinfo  {journal} {New J. Phys.}\ }\textbf {\bibinfo {volume}
  {17}},\ \bibinfo {pages} {075015} (\bibinfo {year} {2015})}\BibitemShut
  {NoStop}%
\bibitem [{\citenamefont {Campisi}\ and\ \citenamefont
  {Fazio}(2016)}]{campisi2016power}%
  \BibitemOpen
  \bibfield  {author} {\bibinfo {author} {\bibfnamefont {M.}~\bibnamefont
  {Campisi}}\ and\ \bibinfo {author} {\bibfnamefont {R.}~\bibnamefont
  {Fazio}},\ }\enquote {\bibinfo {title} {The power of a critical heat
  engine},}\ \href {\doibase 10.1038/ncomms11895} {\bibfield  {journal}
  {\bibinfo  {journal} {Nat. Commun.}\ }\textbf {\bibinfo {volume} {7}},\
  \bibinfo {pages} {11895} (\bibinfo {year} {2016})}\BibitemShut {NoStop}%
\bibitem [{\citenamefont {Jaramillo}\ \emph {et~al.}(2016)\citenamefont
  {Jaramillo}, \citenamefont {Beau},\ and\ \citenamefont {del
  Campo}}]{jaramillo2016quantum}%
  \BibitemOpen
  \bibfield  {author} {\bibinfo {author} {\bibfnamefont {J.}~\bibnamefont
  {Jaramillo}}, \bibinfo {author} {\bibfnamefont {M.}~\bibnamefont {Beau}}, \
  and\ \bibinfo {author} {\bibfnamefont {A.}~\bibnamefont {del Campo}},\
  }\enquote {\bibinfo {title} {Quantum supremacy of many-particle thermal
  machines},}\ \href {\doibase 10.1088/1367-2630/18/7/075019} {\bibfield
  {journal} {\bibinfo  {journal} {New J. Phys.}\ }\textbf {\bibinfo {volume}
  {18}},\ \bibinfo {pages} {075019} (\bibinfo {year} {2016})}\BibitemShut
  {NoStop}%
\bibitem [{\citenamefont {Campaioli}\ \emph {et~al.}(2017)\citenamefont
  {Campaioli}, \citenamefont {Pollock}, \citenamefont {Binder}, \citenamefont
  {C\'eleri}, \citenamefont {Goold}, \citenamefont {Vinjanampathy},\ and\
  \citenamefont {Modi}}]{campaioli2017enhancing}%
  \BibitemOpen
  \bibfield  {author} {\bibinfo {author} {\bibfnamefont {F.}~\bibnamefont
  {Campaioli}}, \bibinfo {author} {\bibfnamefont {F.~A.}\ \bibnamefont
  {Pollock}}, \bibinfo {author} {\bibfnamefont {F.~C.}\ \bibnamefont {Binder}},
  \bibinfo {author} {\bibfnamefont {L.}~\bibnamefont {C\'eleri}}, \bibinfo
  {author} {\bibfnamefont {J.}~\bibnamefont {Goold}}, \bibinfo {author}
  {\bibfnamefont {S.}~\bibnamefont {Vinjanampathy}}, \ and\ \bibinfo {author}
  {\bibfnamefont {K.}~\bibnamefont {Modi}},\ }\enquote {\bibinfo {title}
  {Enhancing the Charging Power of Quantum Batteries},}\ \href {\doibase
  10.1103/PhysRevLett.118.150601} {\bibfield  {journal} {\bibinfo  {journal}
  {Phys. Rev. Lett.}\ }\textbf {\bibinfo {volume} {118}},\ \bibinfo {pages}
  {150601} (\bibinfo {year} {2017})}\BibitemShut {NoStop}%
\bibitem [{\citenamefont {Vroylandt}\ \emph {et~al.}(2017)\citenamefont
  {Vroylandt}, \citenamefont {Esposito},\ and\ \citenamefont
  {Verley}}]{vroylandt2017collective}%
  \BibitemOpen
  \bibfield  {author} {\bibinfo {author} {\bibfnamefont {H.}~\bibnamefont
  {Vroylandt}}, \bibinfo {author} {\bibfnamefont {M.}~\bibnamefont {Esposito}},
  \ and\ \bibinfo {author} {\bibfnamefont {G.}~\bibnamefont {Verley}},\
  }\enquote {\bibinfo {title} {Collective effects enhancing power and
  efficiency},}\ \href {\doibase 10.1209/0295-5075/120/30009} {\bibfield
  {journal} {\bibinfo  {journal} {EPL (Europhys. Lett.)}\ }\textbf {\bibinfo
  {volume} {120}},\ \bibinfo {pages} {30009} (\bibinfo {year}
  {2017})}\BibitemShut {NoStop}%
\bibitem [{\citenamefont {Ferraro}\ \emph {et~al.}(2018)\citenamefont
  {Ferraro}, \citenamefont {Campisi}, \citenamefont {Andolina}, \citenamefont
  {Pellegrini},\ and\ \citenamefont {Polini}}]{ferraro2018high}%
  \BibitemOpen
  \bibfield  {author} {\bibinfo {author} {\bibfnamefont {D.}~\bibnamefont
  {Ferraro}}, \bibinfo {author} {\bibfnamefont {M.}~\bibnamefont {Campisi}},
  \bibinfo {author} {\bibfnamefont {G.~M.}\ \bibnamefont {Andolina}}, \bibinfo
  {author} {\bibfnamefont {V.}~\bibnamefont {Pellegrini}}, \ and\ \bibinfo
  {author} {\bibfnamefont {M.}~\bibnamefont {Polini}},\ }\enquote {\bibinfo
  {title} {High-Power Collective Charging of a Solid-State Quantum Battery},}\
  \href {\doibase 10.1103/PhysRevLett.120.117702} {\bibfield  {journal}
  {\bibinfo  {journal} {Phys. Rev. Lett.}\ }\textbf {\bibinfo {volume} {120}},\
  \bibinfo {pages} {117702} (\bibinfo {year} {2018})}\BibitemShut {NoStop}%
\bibitem [{\citenamefont {Hardal}\ \emph {et~al.}(2018)\citenamefont {Hardal},
  \citenamefont {Paternostro},\ and\ \citenamefont {M\"ustecapl\ifmmode \imath
  \else \i \fi{}o\ifmmode~\breve{g}\else \u{g}\fi{}lu}}]{hardal2018phase}%
  \BibitemOpen
  \bibfield  {author} {\bibinfo {author} {\bibfnamefont {A.~{\"U}.~C.}\
  \bibnamefont {Hardal}}, \bibinfo {author} {\bibfnamefont {M.}~\bibnamefont
  {Paternostro}}, \ and\ \bibinfo {author} {\bibfnamefont {{\"O}.~E.}\
  \bibnamefont {M\"ustecapl\ifmmode \imath \else \i
  \fi{}o\ifmmode~\breve{g}\else \u{g}\fi{}lu}},\ }\enquote {\bibinfo {title}
  {Phase-space interference in extensive and nonextensive quantum heat
  engines},}\ \href {\doibase 10.1103/PhysRevE.97.042127} {\bibfield  {journal}
  {\bibinfo  {journal} {Phys. Rev. E}\ }\textbf {\bibinfo {volume} {97}},\
  \bibinfo {pages} {042127} (\bibinfo {year} {2018})}\BibitemShut {NoStop}%
\bibitem [{\citenamefont {Le}\ \emph {et~al.}(2018)\citenamefont {Le},
  \citenamefont {Levinsen}, \citenamefont {Modi}, \citenamefont {Parish},\ and\
  \citenamefont {Pollock}}]{le2018spin}%
  \BibitemOpen
  \bibfield  {author} {\bibinfo {author} {\bibfnamefont {T.~P.}\ \bibnamefont
  {Le}}, \bibinfo {author} {\bibfnamefont {J.}~\bibnamefont {Levinsen}},
  \bibinfo {author} {\bibfnamefont {K.}~\bibnamefont {Modi}}, \bibinfo {author}
  {\bibfnamefont {M.~M.}\ \bibnamefont {Parish}}, \ and\ \bibinfo {author}
  {\bibfnamefont {F.~A.}\ \bibnamefont {Pollock}},\ }\enquote {\bibinfo {title}
  {Spin-chain model of a many-body quantum battery},}\ \href {\doibase
  10.1103/PhysRevA.97.022106} {\bibfield  {journal} {\bibinfo  {journal} {Phys.
  Rev. A}\ }\textbf {\bibinfo {volume} {97}},\ \bibinfo {pages} {022106}
  (\bibinfo {year} {2018})}\BibitemShut {NoStop}%
\bibitem [{\citenamefont {Dicke}(1954)}]{dicke1954coherence}%
  \BibitemOpen
  \bibfield  {author} {\bibinfo {author} {\bibfnamefont {R.~H.}\ \bibnamefont
  {Dicke}},\ }\enquote {\bibinfo {title} {Coherence in Spontaneous Radiation
  Processes},}\ \href {\doibase 10.1103/PhysRev.93.99} {\bibfield  {journal}
  {\bibinfo  {journal} {Phys. Rev.}\ }\textbf {\bibinfo {volume} {93}},\
  \bibinfo {pages} {99} (\bibinfo {year} {1954})}\BibitemShut {NoStop}%
\bibitem [{\citenamefont {Agarwal}(1970)}]{agarwal1970master}%
  \BibitemOpen
  \bibfield  {author} {\bibinfo {author} {\bibfnamefont {G.~S.}\ \bibnamefont
  {Agarwal}},\ }\enquote {\bibinfo {title} {Master-Equation Approach to
  Spontaneous Emission},}\ \href {\doibase 10.1103/PhysRevA.2.2038} {\bibfield
  {journal} {\bibinfo  {journal} {Phys. Rev. A}\ }\textbf {\bibinfo {volume}
  {2}},\ \bibinfo {pages} {2038} (\bibinfo {year} {1970})}\BibitemShut
  {NoStop}%
\bibitem [{\citenamefont
  {Lehmberg}(1970{\natexlab{a}})}]{lehmberg1970radiation}%
  \BibitemOpen
  \bibfield  {author} {\bibinfo {author} {\bibfnamefont {R.~H.}\ \bibnamefont
  {Lehmberg}},\ }\enquote {\bibinfo {title} {Radiation from an $N$-Atom System.
  I. General Formalism},}\ \href {\doibase 10.1103/PhysRevA.2.883} {\bibfield
  {journal} {\bibinfo  {journal} {Phys. Rev. A}\ }\textbf {\bibinfo {volume}
  {2}},\ \bibinfo {pages} {883} (\bibinfo {year}
  {1970}{\natexlab{a}})}\BibitemShut {NoStop}%
\bibitem [{\citenamefont
  {Lehmberg}(1970{\natexlab{b}})}]{lehmberg1970radiation2}%
  \BibitemOpen
  \bibfield  {author} {\bibinfo {author} {\bibfnamefont {R.~H.}\ \bibnamefont
  {Lehmberg}},\ }\enquote {\bibinfo {title} {Radiation from an $N$-Atom System.
  II. Spontaneous Emission from a Pair of Atoms},}\ \href {\doibase
  10.1103/PhysRevA.2.889} {\bibfield  {journal} {\bibinfo  {journal} {Phys.
  Rev. A}\ }\textbf {\bibinfo {volume} {2}},\ \bibinfo {pages} {889} (\bibinfo
  {year} {1970}{\natexlab{b}})}\BibitemShut {NoStop}%
\bibitem [{\citenamefont {Skribanowitz}\ \emph {et~al.}(1973)\citenamefont
  {Skribanowitz}, \citenamefont {Herman}, \citenamefont {MacGillivray},\ and\
  \citenamefont {Feld}}]{skribanowitz1973observation}%
  \BibitemOpen
  \bibfield  {author} {\bibinfo {author} {\bibfnamefont {N.}~\bibnamefont
  {Skribanowitz}}, \bibinfo {author} {\bibfnamefont {I.~P.}\ \bibnamefont
  {Herman}}, \bibinfo {author} {\bibfnamefont {J.~C.}\ \bibnamefont
  {MacGillivray}}, \ and\ \bibinfo {author} {\bibfnamefont {M.~S.}\
  \bibnamefont {Feld}},\ }\enquote {\bibinfo {title} {Observation of Dicke
  Superradiance in Optically Pumped HF Gas},}\ \href {\doibase
  10.1103/PhysRevLett.30.309} {\bibfield  {journal} {\bibinfo  {journal} {Phys.
  Rev. Lett.}\ }\textbf {\bibinfo {volume} {30}},\ \bibinfo {pages} {309}
  (\bibinfo {year} {1973})}\BibitemShut {NoStop}%
\bibitem [{\citenamefont {Carmichael}(1980)}]{carmichael1980analytical}%
  \BibitemOpen
  \bibfield  {author} {\bibinfo {author} {\bibfnamefont {H.~J.}\ \bibnamefont
  {Carmichael}},\ }\enquote {\bibinfo {title} {Analytical and numerical results
  for the steady state in cooperative resonance fluorescence},}\ \href
  {\doibase 10.1088/0022-3700/13/18/009} {\bibfield  {journal} {\bibinfo
  {journal} {J. Phys. B: At. Mol. Phys.}\ }\textbf {\bibinfo {volume} {13}},\
  \bibinfo {pages} {3551} (\bibinfo {year} {1980})}\BibitemShut {NoStop}%
\bibitem [{\citenamefont {Gross}\ and\ \citenamefont
  {Haroche}(1982)}]{gross1982superradiance}%
  \BibitemOpen
  \bibfield  {author} {\bibinfo {author} {\bibfnamefont {M.}~\bibnamefont
  {Gross}}\ and\ \bibinfo {author} {\bibfnamefont {S.}~\bibnamefont
  {Haroche}},\ }\enquote {\bibinfo {title} {Superradiance: An essay on the
  theory of collective spontaneous emission},}\ \href {\doibase
  10.1016/0370-1573(82)90102-8} {\bibfield  {journal} {\bibinfo  {journal}
  {Phys. Rep.}\ }\textbf {\bibinfo {volume} {93}},\ \bibinfo {pages} {301 }
  (\bibinfo {year} {1982})}\BibitemShut {NoStop}%
\bibitem [{\citenamefont {DeVoe}\ and\ \citenamefont
  {Brewer}(1996)}]{devoe1996observation}%
  \BibitemOpen
  \bibfield  {author} {\bibinfo {author} {\bibfnamefont {R.~G.}\ \bibnamefont
  {DeVoe}}\ and\ \bibinfo {author} {\bibfnamefont {R.~G.}\ \bibnamefont
  {Brewer}},\ }\enquote {\bibinfo {title} {Observation of Superradiant and
  Subradiant Spontaneous Emission of Two Trapped Ions},}\ \href {\doibase
  10.1103/PhysRevLett.76.2049} {\bibfield  {journal} {\bibinfo  {journal}
  {Phys. Rev. Lett.}\ }\textbf {\bibinfo {volume} {76}},\ \bibinfo {pages}
  {2049} (\bibinfo {year} {1996})}\BibitemShut {NoStop}%
\bibitem [{\citenamefont {Hald}\ \emph {et~al.}(1999)\citenamefont {Hald},
  \citenamefont {S\o{}rensen}, \citenamefont {Schori},\ and\ \citenamefont
  {Polzik}}]{hald1999spin}%
  \BibitemOpen
  \bibfield  {author} {\bibinfo {author} {\bibfnamefont {J.}~\bibnamefont
  {Hald}}, \bibinfo {author} {\bibfnamefont {J.~L.}\ \bibnamefont
  {S\o{}rensen}}, \bibinfo {author} {\bibfnamefont {C.}~\bibnamefont {Schori}},
  \ and\ \bibinfo {author} {\bibfnamefont {E.~S.}\ \bibnamefont {Polzik}},\
  }\enquote {\bibinfo {title} {Spin Squeezed Atoms: A Macroscopic Entangled
  Ensemble Created by Light},}\ \href {\doibase 10.1103/PhysRevLett.83.1319}
  {\bibfield  {journal} {\bibinfo  {journal} {Phys. Rev. Lett.}\ }\textbf
  {\bibinfo {volume} {83}},\ \bibinfo {pages} {1319} (\bibinfo {year}
  {1999})}\BibitemShut {NoStop}%
\bibitem [{\citenamefont {Wang}\ \emph {et~al.}(2007)\citenamefont {Wang},
  \citenamefont {Yelin}, \citenamefont {C\^ot\'e}, \citenamefont {Eyler},
  \citenamefont {Farooqi}, \citenamefont {Gould}, \citenamefont
  {Ko\ifmmode~\check{s}\else \v{s}\fi{}trun}, \citenamefont {Tong},\ and\
  \citenamefont {Vrinceanu}}]{wang2007superradiance}%
  \BibitemOpen
  \bibfield  {author} {\bibinfo {author} {\bibfnamefont {T.}~\bibnamefont
  {Wang}}, \bibinfo {author} {\bibfnamefont {S.~F.}\ \bibnamefont {Yelin}},
  \bibinfo {author} {\bibfnamefont {R.}~\bibnamefont {C\^ot\'e}}, \bibinfo
  {author} {\bibfnamefont {E.~E.}\ \bibnamefont {Eyler}}, \bibinfo {author}
  {\bibfnamefont {S.~M.}\ \bibnamefont {Farooqi}}, \bibinfo {author}
  {\bibfnamefont {P.~L.}\ \bibnamefont {Gould}}, \bibinfo {author}
  {\bibfnamefont {M.}~\bibnamefont {Ko\ifmmode~\check{s}\else \v{s}\fi{}trun}},
  \bibinfo {author} {\bibfnamefont {D.}~\bibnamefont {Tong}}, \ and\ \bibinfo
  {author} {\bibfnamefont {D.}~\bibnamefont {Vrinceanu}},\ }\enquote {\bibinfo
  {title} {Superradiance in ultracold Rydberg gases},}\ \href {\doibase
  10.1103/PhysRevA.75.033802} {\bibfield  {journal} {\bibinfo  {journal} {Phys.
  Rev. A}\ }\textbf {\bibinfo {volume} {75}},\ \bibinfo {pages} {033802}
  (\bibinfo {year} {2007})}\BibitemShut {NoStop}%
\bibitem [{\citenamefont {Akkermans}\ \emph {et~al.}(2008)\citenamefont
  {Akkermans}, \citenamefont {Gero},\ and\ \citenamefont
  {Kaiser}}]{akkermans2008photon}%
  \BibitemOpen
  \bibfield  {author} {\bibinfo {author} {\bibfnamefont {E.}~\bibnamefont
  {Akkermans}}, \bibinfo {author} {\bibfnamefont {A.}~\bibnamefont {Gero}}, \
  and\ \bibinfo {author} {\bibfnamefont {R.}~\bibnamefont {Kaiser}},\ }\enquote
  {\bibinfo {title} {Photon Localization and Dicke Superradiance in Atomic
  Gases},}\ \href {\doibase 10.1103/PhysRevLett.101.103602} {\bibfield
  {journal} {\bibinfo  {journal} {Phys. Rev. Lett.}\ }\textbf {\bibinfo
  {volume} {101}},\ \bibinfo {pages} {103602} (\bibinfo {year}
  {2008})}\BibitemShut {NoStop}%
\bibitem [{\citenamefont {Chang}\ \emph {et~al.}(2013)\citenamefont {Chang},
  \citenamefont {Cirac},\ and\ \citenamefont {Kimble}}]{chang2013self}%
  \BibitemOpen
  \bibfield  {author} {\bibinfo {author} {\bibfnamefont {D.~E.}\ \bibnamefont
  {Chang}}, \bibinfo {author} {\bibfnamefont {J.~I.}\ \bibnamefont {Cirac}}, \
  and\ \bibinfo {author} {\bibfnamefont {H.~J.}\ \bibnamefont {Kimble}},\
  }\enquote {\bibinfo {title} {Self-Organization of Atoms along a Nanophotonic
  Waveguide},}\ \href {\doibase 10.1103/PhysRevLett.110.113606} {\bibfield
  {journal} {\bibinfo  {journal} {Phys. Rev. Lett.}\ }\textbf {\bibinfo
  {volume} {110}},\ \bibinfo {pages} {113606} (\bibinfo {year}
  {2013})}\BibitemShut {NoStop}%
\bibitem [{\citenamefont {Ritsch}\ \emph {et~al.}(2013)\citenamefont {Ritsch},
  \citenamefont {Domokos}, \citenamefont {Brennecke},\ and\ \citenamefont
  {Esslinger}}]{ritsch2013cold}%
  \BibitemOpen
  \bibfield  {author} {\bibinfo {author} {\bibfnamefont {H.}~\bibnamefont
  {Ritsch}}, \bibinfo {author} {\bibfnamefont {P.}~\bibnamefont {Domokos}},
  \bibinfo {author} {\bibfnamefont {F.}~\bibnamefont {Brennecke}}, \ and\
  \bibinfo {author} {\bibfnamefont {T.}~\bibnamefont {Esslinger}},\ }\enquote
  {\bibinfo {title} {Cold atoms in cavity-generated dynamical optical
  potentials},}\ \href {\doibase 10.1103/RevModPhys.85.553} {\bibfield
  {journal} {\bibinfo  {journal} {Rev. Mod. Phys.}\ }\textbf {\bibinfo {volume}
  {85}},\ \bibinfo {pages} {553} (\bibinfo {year} {2013})}\BibitemShut
  {NoStop}%
\bibitem [{\citenamefont {van Loo}\ \emph {et~al.}(2013)\citenamefont {van
  Loo}, \citenamefont {Fedorov}, \citenamefont {Lalumi\`ere}, \citenamefont
  {Sanders}, \citenamefont {Blais},\ and\ \citenamefont
  {Wallraff}}]{vanloo2013photon}%
  \BibitemOpen
  \bibfield  {author} {\bibinfo {author} {\bibfnamefont {A.~F.}\ \bibnamefont
  {van Loo}}, \bibinfo {author} {\bibfnamefont {A.}~\bibnamefont {Fedorov}},
  \bibinfo {author} {\bibfnamefont {K.}~\bibnamefont {Lalumi\`ere}}, \bibinfo
  {author} {\bibfnamefont {B.~C.}\ \bibnamefont {Sanders}}, \bibinfo {author}
  {\bibfnamefont {A.}~\bibnamefont {Blais}}, \ and\ \bibinfo {author}
  {\bibfnamefont {A.}~\bibnamefont {Wallraff}},\ }\enquote {\bibinfo {title}
  {Photon-Mediated Interactions Between Distant Artificial Atoms},}\ \href
  {\doibase 10.1126/science.1244324} {\bibfield  {journal} {\bibinfo  {journal}
  {Science}\ }\textbf {\bibinfo {volume} {342}},\ \bibinfo {pages} {1494}
  (\bibinfo {year} {2013})}\BibitemShut {NoStop}%
\bibitem [{\citenamefont {Wickenbrock}\ \emph {et~al.}(2013)\citenamefont
  {Wickenbrock}, \citenamefont {Hemmerling}, \citenamefont {Robb},
  \citenamefont {Emary},\ and\ \citenamefont
  {Renzoni}}]{wickenbrock2013collective}%
  \BibitemOpen
  \bibfield  {author} {\bibinfo {author} {\bibfnamefont {A.}~\bibnamefont
  {Wickenbrock}}, \bibinfo {author} {\bibfnamefont {M.}~\bibnamefont
  {Hemmerling}}, \bibinfo {author} {\bibfnamefont {G.~R.~M.}\ \bibnamefont
  {Robb}}, \bibinfo {author} {\bibfnamefont {C.}~\bibnamefont {Emary}}, \ and\
  \bibinfo {author} {\bibfnamefont {F.}~\bibnamefont {Renzoni}},\ }\enquote
  {\bibinfo {title} {Collective strong coupling in multimode cavity QED},}\
  \href {\doibase 10.1103/PhysRevA.87.043817} {\bibfield  {journal} {\bibinfo
  {journal} {Phys. Rev. A}\ }\textbf {\bibinfo {volume} {87}},\ \bibinfo
  {pages} {043817} (\bibinfo {year} {2013})}\BibitemShut {NoStop}%
\bibitem [{\citenamefont {Maier}\ \emph {et~al.}(2014)\citenamefont {Maier},
  \citenamefont {Kraemer}, \citenamefont {Ostermann},\ and\ \citenamefont
  {Ritsch}}]{maier2014superradiance}%
  \BibitemOpen
  \bibfield  {author} {\bibinfo {author} {\bibfnamefont {T.}~\bibnamefont
  {Maier}}, \bibinfo {author} {\bibfnamefont {S.}~\bibnamefont {Kraemer}},
  \bibinfo {author} {\bibfnamefont {L.}~\bibnamefont {Ostermann}}, \ and\
  \bibinfo {author} {\bibfnamefont {H.}~\bibnamefont {Ritsch}},\ }\enquote
  {\bibinfo {title} {A superradiant clock laser on a magic wavelength optical
  lattice},}\ \href {\doibase 10.1364/OE.22.013269} {\bibfield  {journal}
  {\bibinfo  {journal} {Opt. Express}\ }\textbf {\bibinfo {volume} {22}},\
  \bibinfo {pages} {13269} (\bibinfo {year} {2014})}\BibitemShut {NoStop}%
\bibitem [{\citenamefont {Meir}\ \emph {et~al.}(2014)\citenamefont {Meir},
  \citenamefont {Schwartz}, \citenamefont {Shahmoon}, \citenamefont {Oron},\
  and\ \citenamefont {Ozeri}}]{meir2014cooperative}%
  \BibitemOpen
  \bibfield  {author} {\bibinfo {author} {\bibfnamefont {Z.}~\bibnamefont
  {Meir}}, \bibinfo {author} {\bibfnamefont {O.}~\bibnamefont {Schwartz}},
  \bibinfo {author} {\bibfnamefont {E.}~\bibnamefont {Shahmoon}}, \bibinfo
  {author} {\bibfnamefont {D.}~\bibnamefont {Oron}}, \ and\ \bibinfo {author}
  {\bibfnamefont {R.}~\bibnamefont {Ozeri}},\ }\enquote {\bibinfo {title}
  {Cooperative Lamb Shift in a Mesoscopic Atomic Array},}\ \href {\doibase
  10.1103/PhysRevLett.113.193002} {\bibfield  {journal} {\bibinfo  {journal}
  {Phys. Rev. Lett.}\ }\textbf {\bibinfo {volume} {113}},\ \bibinfo {pages}
  {193002} (\bibinfo {year} {2014})}\BibitemShut {NoStop}%
\bibitem [{\citenamefont {Mlynek}\ \emph {et~al.}(2014)\citenamefont {Mlynek},
  \citenamefont {Abdumalikov}, \citenamefont {Eichler},\ and\ \citenamefont
  {Wallraff}}]{mlynek2014observation}%
  \BibitemOpen
  \bibfield  {author} {\bibinfo {author} {\bibfnamefont {J.~A.}\ \bibnamefont
  {Mlynek}}, \bibinfo {author} {\bibfnamefont {A.~A.}\ \bibnamefont
  {Abdumalikov}}, \bibinfo {author} {\bibfnamefont {C.}~\bibnamefont
  {Eichler}}, \ and\ \bibinfo {author} {\bibfnamefont {A.}~\bibnamefont
  {Wallraff}},\ }\enquote {\bibinfo {title} {Observation of Dicke superradiance
  for two artificial atoms in a cavity with high decay rate},}\ \href {\doibase
  10.1038/ncomms6186} {\bibfield  {journal} {\bibinfo  {journal} {Nat.
  Commun.}\ }\textbf {\bibinfo {volume} {5}},\ \bibinfo {pages} {5186}
  (\bibinfo {year} {2014})}\BibitemShut {NoStop}%
\bibitem [{\citenamefont {Zou}\ \emph {et~al.}(2014)\citenamefont {Zou},
  \citenamefont {Marcos}, \citenamefont {Diehl}, \citenamefont {Putz},
  \citenamefont {Schmiedmayer}, \citenamefont {Majer},\ and\ \citenamefont
  {Rabl}}]{zou2014implementation}%
  \BibitemOpen
  \bibfield  {author} {\bibinfo {author} {\bibfnamefont {L.~J.}\ \bibnamefont
  {Zou}}, \bibinfo {author} {\bibfnamefont {D.}~\bibnamefont {Marcos}},
  \bibinfo {author} {\bibfnamefont {S.}~\bibnamefont {Diehl}}, \bibinfo
  {author} {\bibfnamefont {S.}~\bibnamefont {Putz}}, \bibinfo {author}
  {\bibfnamefont {J.}~\bibnamefont {Schmiedmayer}}, \bibinfo {author}
  {\bibfnamefont {J.}~\bibnamefont {Majer}}, \ and\ \bibinfo {author}
  {\bibfnamefont {P.}~\bibnamefont {Rabl}},\ }\enquote {\bibinfo {title}
  {Implementation of the Dicke Lattice Model in Hybrid Quantum System
  Arrays},}\ \href {\doibase 10.1103/PhysRevLett.113.023603} {\bibfield
  {journal} {\bibinfo  {journal} {Phys. Rev. Lett.}\ }\textbf {\bibinfo
  {volume} {113}},\ \bibinfo {pages} {023603} (\bibinfo {year}
  {2014})}\BibitemShut {NoStop}%
\bibitem [{\citenamefont {Goban}\ \emph {et~al.}(2015)\citenamefont {Goban},
  \citenamefont {Hung}, \citenamefont {Hood}, \citenamefont {Yu}, \citenamefont
  {Muniz}, \citenamefont {Painter},\ and\ \citenamefont
  {Kimble}}]{goban2015superradiance}%
  \BibitemOpen
  \bibfield  {author} {\bibinfo {author} {\bibfnamefont {A.}~\bibnamefont
  {Goban}}, \bibinfo {author} {\bibfnamefont {C.-L.}\ \bibnamefont {Hung}},
  \bibinfo {author} {\bibfnamefont {J.~D.}\ \bibnamefont {Hood}}, \bibinfo
  {author} {\bibfnamefont {S.-P.}\ \bibnamefont {Yu}}, \bibinfo {author}
  {\bibfnamefont {J.~A.}\ \bibnamefont {Muniz}}, \bibinfo {author}
  {\bibfnamefont {O.}~\bibnamefont {Painter}}, \ and\ \bibinfo {author}
  {\bibfnamefont {H.~J.}\ \bibnamefont {Kimble}},\ }\enquote {\bibinfo {title}
  {Superradiance for Atoms Trapped along a Photonic Crystal Waveguide},}\ \href
  {\doibase 10.1103/PhysRevLett.115.063601} {\bibfield  {journal} {\bibinfo
  {journal} {Phys. Rev. Lett.}\ }\textbf {\bibinfo {volume} {115}},\ \bibinfo
  {pages} {063601} (\bibinfo {year} {2015})}\BibitemShut {NoStop}%
\bibitem [{\citenamefont {Klinder}\ \emph {et~al.}(2015)\citenamefont
  {Klinder}, \citenamefont {Ke\ss{}ler}, \citenamefont {Bakhtiari},
  \citenamefont {Thorwart},\ and\ \citenamefont
  {Hemmerich}}]{klinder2015observation}%
  \BibitemOpen
  \bibfield  {author} {\bibinfo {author} {\bibfnamefont {J.}~\bibnamefont
  {Klinder}}, \bibinfo {author} {\bibfnamefont {H.}~\bibnamefont {Ke\ss{}ler}},
  \bibinfo {author} {\bibfnamefont {M.~R.}\ \bibnamefont {Bakhtiari}}, \bibinfo
  {author} {\bibfnamefont {M.}~\bibnamefont {Thorwart}}, \ and\ \bibinfo
  {author} {\bibfnamefont {A.}~\bibnamefont {Hemmerich}},\ }\enquote {\bibinfo
  {title} {Observation of a Superradiant Mott Insulator in the Dicke-Hubbard
  Model},}\ \href {\doibase 10.1103/PhysRevLett.115.230403} {\bibfield
  {journal} {\bibinfo  {journal} {Phys. Rev. Lett.}\ }\textbf {\bibinfo
  {volume} {115}},\ \bibinfo {pages} {230403} (\bibinfo {year}
  {2015})}\BibitemShut {NoStop}%
\bibitem [{\citenamefont {Ara\'ujo}\ \emph {et~al.}(2016)\citenamefont
  {Ara\'ujo}, \citenamefont {Kre\ifmmode \check{s}\else
  \v{s}\fi{}i\ifmmode~\acute{c}\else \'{c}\fi{}}, \citenamefont {Kaiser},\ and\
  \citenamefont {Guerin}}]{araujo2016superradiance}%
  \BibitemOpen
  \bibfield  {author} {\bibinfo {author} {\bibfnamefont {M.~O.}\ \bibnamefont
  {Ara\'ujo}}, \bibinfo {author} {\bibfnamefont {I.}~\bibnamefont {Kre\ifmmode
  \check{s}\else \v{s}\fi{}i\ifmmode~\acute{c}\else \'{c}\fi{}}}, \bibinfo
  {author} {\bibfnamefont {R.}~\bibnamefont {Kaiser}}, \ and\ \bibinfo {author}
  {\bibfnamefont {W.}~\bibnamefont {Guerin}},\ }\enquote {\bibinfo {title}
  {Superradiance in a Large and Dilute Cloud of Cold Atoms in the Linear-Optics
  Regime},}\ \href {\doibase 10.1103/PhysRevLett.117.073002} {\bibfield
  {journal} {\bibinfo  {journal} {Phys. Rev. Lett.}\ }\textbf {\bibinfo
  {volume} {117}},\ \bibinfo {pages} {073002} (\bibinfo {year}
  {2016})}\BibitemShut {NoStop}%
\bibitem [{\citenamefont {Shammah}\ \emph {et~al.}(2017)\citenamefont
  {Shammah}, \citenamefont {Lambert}, \citenamefont {Nori},\ and\ \citenamefont
  {De~Liberato}}]{shammah2017superradiance}%
  \BibitemOpen
  \bibfield  {author} {\bibinfo {author} {\bibfnamefont {N.}~\bibnamefont
  {Shammah}}, \bibinfo {author} {\bibfnamefont {N.}~\bibnamefont {Lambert}},
  \bibinfo {author} {\bibfnamefont {F.}~\bibnamefont {Nori}}, \ and\ \bibinfo
  {author} {\bibfnamefont {S.}~\bibnamefont {De~Liberato}},\ }\enquote
  {\bibinfo {title} {Superradiance with local phase-breaking effects},}\ \href
  {\doibase 10.1103/PhysRevA.96.023863} {\bibfield  {journal} {\bibinfo
  {journal} {Phys. Rev. A}\ }\textbf {\bibinfo {volume} {96}},\ \bibinfo
  {pages} {023863} (\bibinfo {year} {2017})}\BibitemShut {NoStop}%
\bibitem [{\citenamefont {Angerer}\ \emph {et~al.}(2018)\citenamefont
  {Angerer}, \citenamefont {Streltsov}, \citenamefont {Astner}, \citenamefont
  {Putz}, \citenamefont {Sumiya}, \citenamefont {Onoda}, \citenamefont {Isoya},
  \citenamefont {Munro}, \citenamefont {Nemoto}, \citenamefont {Schmiedmayer},\
  and\ \citenamefont {Majer}}]{angerer2018superradiant}%
  \BibitemOpen
  \bibfield  {author} {\bibinfo {author} {\bibfnamefont {A.}~\bibnamefont
  {Angerer}}, \bibinfo {author} {\bibfnamefont {K.}~\bibnamefont {Streltsov}},
  \bibinfo {author} {\bibfnamefont {T.}~\bibnamefont {Astner}}, \bibinfo
  {author} {\bibfnamefont {S.}~\bibnamefont {Putz}}, \bibinfo {author}
  {\bibfnamefont {H.}~\bibnamefont {Sumiya}}, \bibinfo {author} {\bibfnamefont
  {S.}~\bibnamefont {Onoda}}, \bibinfo {author} {\bibfnamefont
  {J.}~\bibnamefont {Isoya}}, \bibinfo {author} {\bibfnamefont {W.~J.}\
  \bibnamefont {Munro}}, \bibinfo {author} {\bibfnamefont {K.}~\bibnamefont
  {Nemoto}}, \bibinfo {author} {\bibfnamefont {J.}~\bibnamefont
  {Schmiedmayer}}, \ and\ \bibinfo {author} {\bibfnamefont {J.}~\bibnamefont
  {Majer}},\ }\enquote {\bibinfo {title} {Superradiant emission from colour
  centres in diamond},}\ \href {\doibase 10.1038/s41567-018-0269-7} {\bibfield
  {journal} {\bibinfo  {journal} {Nat. Phys.}\ } (\bibinfo {year} {2018}),\
  10.1038/s41567-018-0269-7}\BibitemShut {NoStop}%
\bibitem [{\citenamefont {Mandel}\ and\ \citenamefont
  {Wolf}(1995)}]{mandelbook}%
  \BibitemOpen
  \bibfield  {author} {\bibinfo {author} {\bibfnamefont {L.}~\bibnamefont
  {Mandel}}\ and\ \bibinfo {author} {\bibfnamefont {E.}~\bibnamefont {Wolf}},\
  }\href@noop {} {\emph {\bibinfo {title} {Optical coherence and quantum
  optics}}}\ (\bibinfo  {publisher} {Cambridge University Press},\ \bibinfo
  {address} {Cambridge},\ \bibinfo {year} {1995})\BibitemShut {NoStop}%
\bibitem [{\citenamefont {Breuer}\ and\ \citenamefont
  {Petruccione}(2002)}]{breuerbook}%
  \BibitemOpen
  \bibfield  {author} {\bibinfo {author} {\bibfnamefont {H.-P.}\ \bibnamefont
  {Breuer}}\ and\ \bibinfo {author} {\bibfnamefont {F.}~\bibnamefont
  {Petruccione}},\ }\href@noop {} {\emph {\bibinfo {title} {The Theory of Open
  Quantum Systems}}}\ (\bibinfo  {publisher} {Oxford University Press},\
  \bibinfo {address} {Oxford},\ \bibinfo {year} {2002})\BibitemShut {NoStop}%
\bibitem [{\citenamefont {Shahmoon}\ and\ \citenamefont
  {Kurizki}(2013)}]{shahmoon2013nonradiative}%
  \BibitemOpen
  \bibfield  {author} {\bibinfo {author} {\bibfnamefont {E.}~\bibnamefont
  {Shahmoon}}\ and\ \bibinfo {author} {\bibfnamefont {G.}~\bibnamefont
  {Kurizki}},\ }\enquote {\bibinfo {title} {Nonradiative interaction and
  entanglement between distant atoms},}\ \href {\doibase
  10.1103/PhysRevA.87.033831} {\bibfield  {journal} {\bibinfo  {journal} {Phys.
  Rev. A}\ }\textbf {\bibinfo {volume} {87}},\ \bibinfo {pages} {033831}
  (\bibinfo {year} {2013})}\BibitemShut {NoStop}%
\bibitem [{\citenamefont {Lalumi\`ere}\ \emph {et~al.}(2013)\citenamefont
  {Lalumi\`ere}, \citenamefont {Sanders}, \citenamefont {van Loo},
  \citenamefont {Fedorov}, \citenamefont {Wallraff},\ and\ \citenamefont
  {Blais}}]{lalumiere2013input}%
  \BibitemOpen
  \bibfield  {author} {\bibinfo {author} {\bibfnamefont {K.}~\bibnamefont
  {Lalumi\`ere}}, \bibinfo {author} {\bibfnamefont {B.~C.}\ \bibnamefont
  {Sanders}}, \bibinfo {author} {\bibfnamefont {A.~F.}\ \bibnamefont {van
  Loo}}, \bibinfo {author} {\bibfnamefont {A.}~\bibnamefont {Fedorov}},
  \bibinfo {author} {\bibfnamefont {A.}~\bibnamefont {Wallraff}}, \ and\
  \bibinfo {author} {\bibfnamefont {A.}~\bibnamefont {Blais}},\ }\enquote
  {\bibinfo {title} {Input-output theory for waveguide QED with an ensemble of
  inhomogeneous atoms},}\ \href {\doibase 10.1103/PhysRevA.88.043806}
  {\bibfield  {journal} {\bibinfo  {journal} {Phys. Rev. A}\ }\textbf {\bibinfo
  {volume} {88}},\ \bibinfo {pages} {043806} (\bibinfo {year}
  {2013})}\BibitemShut {NoStop}%
\bibitem [{\citenamefont {Scully}\ \emph {et~al.}(2006)\citenamefont {Scully},
  \citenamefont {Fry}, \citenamefont {Ooi},\ and\ \citenamefont
  {W\'odkiewicz}}]{scully2006directed}%
  \BibitemOpen
  \bibfield  {author} {\bibinfo {author} {\bibfnamefont {M.~O.}\ \bibnamefont
  {Scully}}, \bibinfo {author} {\bibfnamefont {E.~S.}\ \bibnamefont {Fry}},
  \bibinfo {author} {\bibfnamefont {C.~H.~R.}\ \bibnamefont {Ooi}}, \ and\
  \bibinfo {author} {\bibfnamefont {K.}~\bibnamefont {W\'odkiewicz}},\
  }\enquote {\bibinfo {title} {Directed Spontaneous Emission from an Extended
  Ensemble of $N$ Atoms: Timing Is Everything},}\ \href {\doibase
  10.1103/PhysRevLett.96.010501} {\bibfield  {journal} {\bibinfo  {journal}
  {Phys. Rev. Lett.}\ }\textbf {\bibinfo {volume} {96}},\ \bibinfo {pages}
  {010501} (\bibinfo {year} {2006})}\BibitemShut {NoStop}%
\bibitem [{\citenamefont {Mazets}\ and\ \citenamefont
  {Kurizki}(2007)}]{mazets2007multiatom}%
  \BibitemOpen
  \bibfield  {author} {\bibinfo {author} {\bibfnamefont {I.~E.}\ \bibnamefont
  {Mazets}}\ and\ \bibinfo {author} {\bibfnamefont {G.}~\bibnamefont
  {Kurizki}},\ }\enquote {\bibinfo {title} {Multiatom cooperative emission
  following single-photon absorption: Dicke-state dynamics},}\ \href {\doibase
  10.1088/0953-4075/40/6/F01} {\bibfield  {journal} {\bibinfo  {journal} {J.
  Phys. B: At. Mol. Opt. Phys.}\ }\textbf {\bibinfo {volume} {40}},\ \bibinfo
  {pages} {F105} (\bibinfo {year} {2007})}\BibitemShut {NoStop}%
\bibitem [{\citenamefont {Svidzinsky}\ \emph
  {et~al.}(2008{\natexlab{a}})\citenamefont {Svidzinsky}, \citenamefont
  {Chang},\ and\ \citenamefont {Scully}}]{svidzinsky2008dynamical}%
  \BibitemOpen
  \bibfield  {author} {\bibinfo {author} {\bibfnamefont {A.~A.}\ \bibnamefont
  {Svidzinsky}}, \bibinfo {author} {\bibfnamefont {J.-T.}\ \bibnamefont
  {Chang}}, \ and\ \bibinfo {author} {\bibfnamefont {M.~O.}\ \bibnamefont
  {Scully}},\ }\enquote {\bibinfo {title} {Dynamical Evolution of Correlated
  Spontaneous Emission of a Single Photon from a Uniformly Excited Cloud of $N$
  Atoms},}\ \href {\doibase 10.1103/PhysRevLett.100.160504} {\bibfield
  {journal} {\bibinfo  {journal} {Phys. Rev. Lett.}\ }\textbf {\bibinfo
  {volume} {100}},\ \bibinfo {pages} {160504} (\bibinfo {year}
  {2008}{\natexlab{a}})}\BibitemShut {NoStop}%
\bibitem [{\citenamefont {Bonifacio}\ and\ \citenamefont
  {Lugiato}(1975)}]{bonifacio1975cooperative}%
  \BibitemOpen
  \bibfield  {author} {\bibinfo {author} {\bibfnamefont {R.}~\bibnamefont
  {Bonifacio}}\ and\ \bibinfo {author} {\bibfnamefont {L.~A.}\ \bibnamefont
  {Lugiato}},\ }\enquote {\bibinfo {title} {Cooperative radiation processes in
  two-level systems: Superfluorescence},}\ \href {\doibase
  10.1103/PhysRevA.11.1507} {\bibfield  {journal} {\bibinfo  {journal} {Phys.
  Rev. A}\ }\textbf {\bibinfo {volume} {11}},\ \bibinfo {pages} {1507}
  (\bibinfo {year} {1975})}\BibitemShut {NoStop}%
\bibitem [{\citenamefont {Scully}\ and\ \citenamefont
  {Svidzinsky}(2009)}]{scully2009super}%
  \BibitemOpen
  \bibfield  {author} {\bibinfo {author} {\bibfnamefont {M.~O.}\ \bibnamefont
  {Scully}}\ and\ \bibinfo {author} {\bibfnamefont {A.~A.}\ \bibnamefont
  {Svidzinsky}},\ }\enquote {\bibinfo {title} {The Super of Superradiance},}\
  \href {\doibase 10.1126/science.1176695} {\bibfield  {journal} {\bibinfo
  {journal} {Science}\ }\textbf {\bibinfo {volume} {325}},\ \bibinfo {pages}
  {1510} (\bibinfo {year} {2009})}\BibitemShut {NoStop}%
\bibitem [{\citenamefont {Svidzinsky}\ \emph
  {et~al.}(2008{\natexlab{b}})\citenamefont {Svidzinsky}, \citenamefont
  {Chang}, \citenamefont {Lipkin},\ and\ \citenamefont
  {Scully}}]{svidzinsky2008fermi}%
  \BibitemOpen
  \bibfield  {author} {\bibinfo {author} {\bibfnamefont {A.}~\bibnamefont
  {Svidzinsky}}, \bibinfo {author} {\bibfnamefont {J.-T.}\ \bibnamefont
  {Chang}}, \bibinfo {author} {\bibfnamefont {H.}~\bibnamefont {Lipkin}}, \
  and\ \bibinfo {author} {\bibfnamefont {M.}~\bibnamefont {Scully}},\ }\enquote
  {\bibinfo {title} {Fermi's golden rule does not adequately describe Dicke's
  superradiance},}\ \href {\doibase 10.1080/09500340802334579} {\bibfield
  {journal} {\bibinfo  {journal} {J. Mod. Opt.}\ }\textbf {\bibinfo {volume}
  {55}},\ \bibinfo {pages} {3369} (\bibinfo {year}
  {2008}{\natexlab{b}})}\BibitemShut {NoStop}%
\bibitem [{\citenamefont {Svidzinsky}\ \emph {et~al.}(2010)\citenamefont
  {Svidzinsky}, \citenamefont {Chang},\ and\ \citenamefont
  {Scully}}]{svidzinsky2010cooperative}%
  \BibitemOpen
  \bibfield  {author} {\bibinfo {author} {\bibfnamefont {A.~A.}\ \bibnamefont
  {Svidzinsky}}, \bibinfo {author} {\bibfnamefont {J.-T.}\ \bibnamefont
  {Chang}}, \ and\ \bibinfo {author} {\bibfnamefont {M.~O.}\ \bibnamefont
  {Scully}},\ }\enquote {\bibinfo {title} {Cooperative spontaneous emission of
  $N$ atoms: Many-body eigenstates, the effect of virtual Lamb shift processes,
  and analogy with radiation of $N$ classical oscillators},}\ \href {\doibase
  10.1103/PhysRevA.81.053821} {\bibfield  {journal} {\bibinfo  {journal} {Phys.
  Rev. A}\ }\textbf {\bibinfo {volume} {81}},\ \bibinfo {pages} {053821}
  (\bibinfo {year} {2010})}\BibitemShut {NoStop}%
\bibitem [{\citenamefont {Gelbwaser-Klimovsky}\ \emph
  {et~al.}(2013)\citenamefont {Gelbwaser-Klimovsky}, \citenamefont {Alicki},\
  and\ \citenamefont {Kurizki}}]{gelbwaser2013minimal}%
  \BibitemOpen
  \bibfield  {author} {\bibinfo {author} {\bibfnamefont {D.}~\bibnamefont
  {Gelbwaser-Klimovsky}}, \bibinfo {author} {\bibfnamefont {R.}~\bibnamefont
  {Alicki}}, \ and\ \bibinfo {author} {\bibfnamefont {G.}~\bibnamefont
  {Kurizki}},\ }\enquote {\bibinfo {title} {Minimal universal quantum heat
  machine},}\ \href {\doibase 10.1103/PhysRevE.87.012140} {\bibfield  {journal}
  {\bibinfo  {journal} {Phys. Rev. E}\ }\textbf {\bibinfo {volume} {87}},\
  \bibinfo {pages} {012140} (\bibinfo {year} {2013})}\BibitemShut {NoStop}%
\bibitem [{\citenamefont {Alicki}(2014)}]{alicki2014quantum}%
  \BibitemOpen
  \bibfield  {author} {\bibinfo {author} {\bibfnamefont {R.}~\bibnamefont
  {Alicki}},\ }\enquote {\bibinfo {title} {Quantum Thermodynamics: An Example
  of Two-Level Quantum Machine},}\ \href {\doibase 10.1142/S1230161214400022}
  {\bibfield  {journal} {\bibinfo  {journal} {Open Syst. Inf. Dyn.}\ }\textbf
  {\bibinfo {volume} {21}},\ \bibinfo {pages} {1440002} (\bibinfo {year}
  {2014})}\BibitemShut {NoStop}%
\bibitem [{\citenamefont {Walls}\ and\ \citenamefont
  {Milburn}(1994)}]{wallsbook}%
  \BibitemOpen
  \bibfield  {author} {\bibinfo {author} {\bibfnamefont {D.~F.}\ \bibnamefont
  {Walls}}\ and\ \bibinfo {author} {\bibfnamefont {G.~J.}\ \bibnamefont
  {Milburn}},\ }\href@noop {} {\emph {\bibinfo {title} {Quantum Optics}}},\
  \bibinfo {edition} {1st}\ ed.\ (\bibinfo  {publisher} {Springer-Verlag},\
  \bibinfo {address} {Berlin},\ \bibinfo {year} {1994})\BibitemShut {NoStop}%
\bibitem [{\citenamefont {Carmichael}(1993)}]{carmichaelbook}%
  \BibitemOpen
  \bibfield  {author} {\bibinfo {author} {\bibfnamefont {H.}~\bibnamefont
  {Carmichael}},\ }\href@noop {} {\emph {\bibinfo {title} {An Open Systems
  Approach to Quantum Optics}}}\ (\bibinfo  {publisher} {Springer-Verlag},\
  \bibinfo {address} {Berlin Heidelberg},\ \bibinfo {year} {1993})\BibitemShut
  {NoStop}%
\bibitem [{\citenamefont {Tasgin}(2017)}]{tasgin2017many}%
  \BibitemOpen
  \bibfield  {author} {\bibinfo {author} {\bibfnamefont {M.~E.}\ \bibnamefont
  {Tasgin}},\ }\enquote {\bibinfo {title} {Many-Particle Entanglement Criterion
  for Superradiantlike States},}\ \href {\doibase
  10.1103/PhysRevLett.119.033601} {\bibfield  {journal} {\bibinfo  {journal}
  {Phys. Rev. Lett.}\ }\textbf {\bibinfo {volume} {119}},\ \bibinfo {pages}
  {033601} (\bibinfo {year} {2017})}\BibitemShut {NoStop}%
\bibitem [{\citenamefont {Schwabl}(2007)}]{schwablbookqm1}%
  \BibitemOpen
  \bibfield  {author} {\bibinfo {author} {\bibfnamefont {F.}~\bibnamefont
  {Schwabl}},\ }\href@noop {} {\emph {\bibinfo {title} {Quantum Mechanics}}},\
  \bibinfo {edition} {forth}\ ed.\ (\bibinfo  {publisher} {Springer-Verlag},\
  \bibinfo {address} {Berlin Heidelberg},\ \bibinfo {year} {2007})\BibitemShut
  {NoStop}%
\bibitem [{\citenamefont {Zachos}(1992)}]{zachos1992altering}%
  \BibitemOpen
  \bibfield  {author} {\bibinfo {author} {\bibfnamefont {C.~K.}\ \bibnamefont
  {Zachos}},\ }\enquote {\bibinfo {title} {Altering the symmetry of wave
  functions in quantum algebras and supersymmetry},}\ \href {\doibase
  10.1142/S0217732392001270} {\bibfield  {journal} {\bibinfo  {journal} {Mod.
  Phys. Lett. A}\ }\textbf {\bibinfo {volume} {7}},\ \bibinfo {pages} {1595}
  (\bibinfo {year} {1992})}\BibitemShut {NoStop}%
\bibitem [{\citenamefont {Geva}\ \emph {et~al.}(1995)\citenamefont {Geva},
  \citenamefont {Kosloff},\ and\ \citenamefont {Skinner}}]{geva1995relaxation}%
  \BibitemOpen
  \bibfield  {author} {\bibinfo {author} {\bibfnamefont {E.}~\bibnamefont
  {Geva}}, \bibinfo {author} {\bibfnamefont {R.}~\bibnamefont {Kosloff}}, \
  and\ \bibinfo {author} {\bibfnamefont {J.~L.}\ \bibnamefont {Skinner}},\
  }\enquote {\bibinfo {title} {On the relaxation of a two-level system driven
  by a strong electromagnetic field},}\ \href {\doibase 10.1063/1.468844}
  {\bibfield  {journal} {\bibinfo  {journal} {J. Chem. Phys.}\ }\textbf
  {\bibinfo {volume} {102}},\ \bibinfo {pages} {8541} (\bibinfo {year}
  {1995})}\BibitemShut {NoStop}%
\bibitem [{\citenamefont {Alicki}\ \emph {et~al.}(2006)\citenamefont {Alicki},
  \citenamefont {Lidar},\ and\ \citenamefont {Zanardi}}]{alicki2006internal}%
  \BibitemOpen
  \bibfield  {author} {\bibinfo {author} {\bibfnamefont {R.}~\bibnamefont
  {Alicki}}, \bibinfo {author} {\bibfnamefont {D.~A.}\ \bibnamefont {Lidar}}, \
  and\ \bibinfo {author} {\bibfnamefont {P.}~\bibnamefont {Zanardi}},\
  }\enquote {\bibinfo {title} {Internal consistency of fault-tolerant quantum
  error correction in light of rigorous derivations of the quantum Markovian
  limit},}\ \href {\doibase 10.1103/PhysRevA.73.052311} {\bibfield  {journal}
  {\bibinfo  {journal} {Phys. Rev. A}\ }\textbf {\bibinfo {volume} {73}},\
  \bibinfo {pages} {052311} (\bibinfo {year} {2006})}\BibitemShut {NoStop}%
\bibitem [{\citenamefont {Szczygielski}\ \emph {et~al.}(2013)\citenamefont
  {Szczygielski}, \citenamefont {Gelbwaser-Klimovsky},\ and\ \citenamefont
  {Alicki}}]{szczygielski2013markovian}%
  \BibitemOpen
  \bibfield  {author} {\bibinfo {author} {\bibfnamefont {K.}~\bibnamefont
  {Szczygielski}}, \bibinfo {author} {\bibfnamefont {D.}~\bibnamefont
  {Gelbwaser-Klimovsky}}, \ and\ \bibinfo {author} {\bibfnamefont
  {R.}~\bibnamefont {Alicki}},\ }\enquote {\bibinfo {title} {Markovian master
  equation and thermodynamics of a two-level system in a strong laser field},}\
  \href {\doibase 10.1103/PhysRevE.87.012120} {\bibfield  {journal} {\bibinfo
  {journal} {Phys. Rev. E}\ }\textbf {\bibinfo {volume} {87}},\ \bibinfo
  {pages} {012120} (\bibinfo {year} {2013})}\BibitemShut {NoStop}%
\bibitem [{\citenamefont {Szczygielski}(2014)}]{szczygielski2014application}%
  \BibitemOpen
  \bibfield  {author} {\bibinfo {author} {\bibfnamefont {K.}~\bibnamefont
  {Szczygielski}},\ }\enquote {\bibinfo {title} {On the application of Floquet
  theorem in development of time-dependent Lindbladians},}\ \href {\doibase
  10.1063/1.4891401} {\bibfield  {journal} {\bibinfo  {journal} {J. Math.
  Phys.}\ }\textbf {\bibinfo {volume} {55}},\ \bibinfo {eid} {083506} (\bibinfo
  {year} {2014})}\BibitemShut {NoStop}%
\bibitem [{\citenamefont {Brandner}\ and\ \citenamefont
  {Seifert}(2016)}]{brandner2016periodic}%
  \BibitemOpen
  \bibfield  {author} {\bibinfo {author} {\bibfnamefont {K.}~\bibnamefont
  {Brandner}}\ and\ \bibinfo {author} {\bibfnamefont {U.}~\bibnamefont
  {Seifert}},\ }\enquote {\bibinfo {title} {Periodic thermodynamics of open
  quantum systems},}\ \href {\doibase 10.1103/PhysRevE.93.062134} {\bibfield
  {journal} {\bibinfo  {journal} {Phys. Rev. E}\ }\textbf {\bibinfo {volume}
  {93}},\ \bibinfo {pages} {062134} (\bibinfo {year} {2016})}\BibitemShut
  {NoStop}%
\bibitem [{\citenamefont {Meiser}\ and\ \citenamefont
  {Holland}(2010)}]{meiser2010steady}%
  \BibitemOpen
  \bibfield  {author} {\bibinfo {author} {\bibfnamefont {D.}~\bibnamefont
  {Meiser}}\ and\ \bibinfo {author} {\bibfnamefont {M.~J.}\ \bibnamefont
  {Holland}},\ }\enquote {\bibinfo {title} {Steady-state superradiance with
  alkaline-earth-metal atoms},}\ \href {\doibase 10.1103/PhysRevA.81.033847}
  {\bibfield  {journal} {\bibinfo  {journal} {Phys. Rev. A}\ }\textbf {\bibinfo
  {volume} {81}},\ \bibinfo {pages} {033847} (\bibinfo {year}
  {2010})}\BibitemShut {NoStop}%
\bibitem [{\citenamefont {Gonz\'alez-Tudela}\ \emph {et~al.}(2015)\citenamefont
  {Gonz\'alez-Tudela}, \citenamefont {Paulisch}, \citenamefont {Chang},
  \citenamefont {Kimble},\ and\ \citenamefont
  {Cirac}}]{gonzalez2015deterministic}%
  \BibitemOpen
  \bibfield  {author} {\bibinfo {author} {\bibfnamefont {A.}~\bibnamefont
  {Gonz\'alez-Tudela}}, \bibinfo {author} {\bibfnamefont {V.}~\bibnamefont
  {Paulisch}}, \bibinfo {author} {\bibfnamefont {D.~E.}\ \bibnamefont {Chang}},
  \bibinfo {author} {\bibfnamefont {H.~J.}\ \bibnamefont {Kimble}}, \ and\
  \bibinfo {author} {\bibfnamefont {J.~I.}\ \bibnamefont {Cirac}},\ }\enquote
  {\bibinfo {title} {Deterministic Generation of Arbitrary Photonic States
  Assisted by Dissipation},}\ \href {\doibase 10.1103/PhysRevLett.115.163603}
  {\bibfield  {journal} {\bibinfo  {journal} {Phys. Rev. Lett.}\ }\textbf
  {\bibinfo {volume} {115}},\ \bibinfo {pages} {163603} (\bibinfo {year}
  {2015})}\BibitemShut {NoStop}%
\bibitem [{\citenamefont {L{\'e}onard}\ \emph {et~al.}(2017)\citenamefont
  {L{\'e}onard}, \citenamefont {Morales}, \citenamefont {Zupancic},
  \citenamefont {Esslinger},\ and\ \citenamefont
  {Donner}}]{leonard2017supersolid}%
  \BibitemOpen
  \bibfield  {author} {\bibinfo {author} {\bibfnamefont {J.}~\bibnamefont
  {L{\'e}onard}}, \bibinfo {author} {\bibfnamefont {A.}~\bibnamefont
  {Morales}}, \bibinfo {author} {\bibfnamefont {P.}~\bibnamefont {Zupancic}},
  \bibinfo {author} {\bibfnamefont {T.}~\bibnamefont {Esslinger}}, \ and\
  \bibinfo {author} {\bibfnamefont {T.}~\bibnamefont {Donner}},\ }\enquote
  {\bibinfo {title} {Supersolid formation in a quantum gas breaking a
  continuous translational symmetry},}\ \href {\doibase 10.1038/nature21067}
  {\bibfield  {journal} {\bibinfo  {journal} {Nature}\ }\textbf {\bibinfo
  {volume} {543}},\ \bibinfo {pages} {87} (\bibinfo {year} {2017})}\BibitemShut
  {NoStop}%
\bibitem [{\citenamefont {Keller}\ \emph {et~al.}(2018)\citenamefont {Keller},
  \citenamefont {Torggler}, \citenamefont {J\"ager}, \citenamefont {Sch\"utz},
  \citenamefont {Ritsch},\ and\ \citenamefont {Morigi}}]{keller2018quenches}%
  \BibitemOpen
  \bibfield  {author} {\bibinfo {author} {\bibfnamefont {T.}~\bibnamefont
  {Keller}}, \bibinfo {author} {\bibfnamefont {V.}~\bibnamefont {Torggler}},
  \bibinfo {author} {\bibfnamefont {S.~B.}\ \bibnamefont {J\"ager}}, \bibinfo
  {author} {\bibfnamefont {S.}~\bibnamefont {Sch\"utz}}, \bibinfo {author}
  {\bibfnamefont {H.}~\bibnamefont {Ritsch}}, \ and\ \bibinfo {author}
  {\bibfnamefont {G.}~\bibnamefont {Morigi}},\ }\enquote {\bibinfo {title}
  {Quenches across the self-organization transition in multimode cavities},}\
  \href {\doibase 10.1088/1367-2630/aaa161} {\bibfield  {journal} {\bibinfo
  {journal} {New J. Phys.}\ }\textbf {\bibinfo {volume} {20}},\ \bibinfo
  {pages} {025004} (\bibinfo {year} {2018})}\BibitemShut {NoStop}%
\bibitem [{\citenamefont {Kopylov}\ \emph {et~al.}(2015)\citenamefont
  {Kopylov}, \citenamefont {Radonji\ifmmode~\acute{c}\else \'{c}\fi{}},
  \citenamefont {Brandes}, \citenamefont {Bala\ifmmode~\check{z}\else
  \v{z}\fi{}},\ and\ \citenamefont {Pelster}}]{kopylov2015dissipative}%
  \BibitemOpen
  \bibfield  {author} {\bibinfo {author} {\bibfnamefont {W.}~\bibnamefont
  {Kopylov}}, \bibinfo {author} {\bibfnamefont {M.}~\bibnamefont
  {Radonji\ifmmode~\acute{c}\else \'{c}\fi{}}}, \bibinfo {author}
  {\bibfnamefont {T.}~\bibnamefont {Brandes}}, \bibinfo {author} {\bibfnamefont
  {A.}~\bibnamefont {Bala\ifmmode~\check{z}\else \v{z}\fi{}}}, \ and\ \bibinfo
  {author} {\bibfnamefont {A.}~\bibnamefont {Pelster}},\ }\enquote {\bibinfo
  {title} {Dissipative two-mode Tavis-Cummings model with time-delayed feedback
  control},}\ \href {\doibase 10.1103/PhysRevA.92.063832} {\bibfield  {journal}
  {\bibinfo  {journal} {Phys. Rev. A}\ }\textbf {\bibinfo {volume} {92}},\
  \bibinfo {pages} {063832} (\bibinfo {year} {2015})}\BibitemShut {NoStop}%
\bibitem [{\citenamefont {Moodie}\ \emph {et~al.}(2018)\citenamefont {Moodie},
  \citenamefont {Ballantine},\ and\ \citenamefont
  {Keeling}}]{moodie2018generalized}%
  \BibitemOpen
  \bibfield  {author} {\bibinfo {author} {\bibfnamefont {R.~I.}\ \bibnamefont
  {Moodie}}, \bibinfo {author} {\bibfnamefont {K.~E.}\ \bibnamefont
  {Ballantine}}, \ and\ \bibinfo {author} {\bibfnamefont {J.}~\bibnamefont
  {Keeling}},\ }\enquote {\bibinfo {title} {Generalized classes of continuous
  symmetries in two-mode Dicke models},}\ \href {\doibase
  10.1103/PhysRevA.97.033802} {\bibfield  {journal} {\bibinfo  {journal} {Phys.
  Rev. A}\ }\textbf {\bibinfo {volume} {97}},\ \bibinfo {pages} {033802}
  (\bibinfo {year} {2018})}\BibitemShut {NoStop}%
\bibitem [{\citenamefont {Bradac}\ \emph {et~al.}(2017)\citenamefont {Bradac},
  \citenamefont {Johnsson}, \citenamefont {van Breugel}, \citenamefont
  {Baragiola}, \citenamefont {Martin}, \citenamefont {Juan}, \citenamefont
  {Brennen},\ and\ \citenamefont {Volz}}]{bradac2017room}%
  \BibitemOpen
  \bibfield  {author} {\bibinfo {author} {\bibfnamefont {C.}~\bibnamefont
  {Bradac}}, \bibinfo {author} {\bibfnamefont {M.~T.}\ \bibnamefont
  {Johnsson}}, \bibinfo {author} {\bibfnamefont {M.}~\bibnamefont {van
  Breugel}}, \bibinfo {author} {\bibfnamefont {B.~Q.}\ \bibnamefont
  {Baragiola}}, \bibinfo {author} {\bibfnamefont {R.}~\bibnamefont {Martin}},
  \bibinfo {author} {\bibfnamefont {M.~L.}\ \bibnamefont {Juan}}, \bibinfo
  {author} {\bibfnamefont {G.~K.}\ \bibnamefont {Brennen}}, \ and\ \bibinfo
  {author} {\bibfnamefont {T.}~\bibnamefont {Volz}},\ }\enquote {\bibinfo
  {title} {Room-temperature spontaneous superradiance from single diamond
  nanocrystals},}\ \href {\doibase 10.1038/s41467-017-01397-4} {\bibfield
  {journal} {\bibinfo  {journal} {Nat. Commun.}\ }\textbf {\bibinfo {volume}
  {8}},\ \bibinfo {pages} {1205} (\bibinfo {year} {2017})}\BibitemShut
  {NoStop}%
\bibitem [{\citenamefont {Brandner}\ \emph {et~al.}(2017)\citenamefont
  {Brandner}, \citenamefont {Bauer},\ and\ \citenamefont
  {Seifert}}]{brandner2017universal}%
  \BibitemOpen
  \bibfield  {author} {\bibinfo {author} {\bibfnamefont {K.}~\bibnamefont
  {Brandner}}, \bibinfo {author} {\bibfnamefont {M.}~\bibnamefont {Bauer}}, \
  and\ \bibinfo {author} {\bibfnamefont {U.}~\bibnamefont {Seifert}},\
  }\enquote {\bibinfo {title} {Universal Coherence-Induced Power Losses of
  Quantum Heat Engines in Linear Response},}\ \href {\doibase
  10.1103/PhysRevLett.119.170602} {\bibfield  {journal} {\bibinfo  {journal}
  {Phys. Rev. Lett.}\ }\textbf {\bibinfo {volume} {119}},\ \bibinfo {pages}
  {170602} (\bibinfo {year} {2017})}\BibitemShut {NoStop}%
\bibitem [{\citenamefont {Gelbwaser-Klimovsky}\ \emph
  {et~al.}(2015{\natexlab{b}})\citenamefont {Gelbwaser-Klimovsky},
  \citenamefont {Szczygielski}, \citenamefont {Vogl}, \citenamefont {Sa\ss{}},
  \citenamefont {Alicki}, \citenamefont {Kurizki},\ and\ \citenamefont
  {Weitz}}]{gelbwaser2015laser}%
  \BibitemOpen
  \bibfield  {author} {\bibinfo {author} {\bibfnamefont {D.}~\bibnamefont
  {Gelbwaser-Klimovsky}}, \bibinfo {author} {\bibfnamefont {K.}~\bibnamefont
  {Szczygielski}}, \bibinfo {author} {\bibfnamefont {U.}~\bibnamefont {Vogl}},
  \bibinfo {author} {\bibfnamefont {A.}~\bibnamefont {Sa\ss{}}}, \bibinfo
  {author} {\bibfnamefont {R.}~\bibnamefont {Alicki}}, \bibinfo {author}
  {\bibfnamefont {G.}~\bibnamefont {Kurizki}}, \ and\ \bibinfo {author}
  {\bibfnamefont {M.}~\bibnamefont {Weitz}},\ }\enquote {\bibinfo {title}
  {Laser-induced cooling of broadband heat reservoirs},}\ \href {\doibase
  10.1103/PhysRevA.91.023431} {\bibfield  {journal} {\bibinfo  {journal} {Phys.
  Rev. A}\ }\textbf {\bibinfo {volume} {91}},\ \bibinfo {pages} {023431}
  (\bibinfo {year} {2015}{\natexlab{b}})}\BibitemShut {NoStop}%
\bibitem [{\citenamefont {Prasanna~Venkatesh}\ \emph
  {et~al.}(2018)\citenamefont {Prasanna~Venkatesh}, \citenamefont {Juan},\ and\
  \citenamefont {Romero-Isart}}]{venkatesh2018cooperative}%
  \BibitemOpen
  \bibfield  {author} {\bibinfo {author} {\bibfnamefont {B.}~\bibnamefont
  {Prasanna~Venkatesh}}, \bibinfo {author} {\bibfnamefont {M.~L.}\ \bibnamefont
  {Juan}}, \ and\ \bibinfo {author} {\bibfnamefont {O.}~\bibnamefont
  {Romero-Isart}},\ }\enquote {\bibinfo {title} {Cooperative Effects in Closely
  Packed Quantum Emitters with Collective Dephasing},}\ \href {\doibase
  10.1103/PhysRevLett.120.033602} {\bibfield  {journal} {\bibinfo  {journal}
  {Phys. Rev. Lett.}\ }\textbf {\bibinfo {volume} {120}},\ \bibinfo {pages}
  {033602} (\bibinfo {year} {2018})}\BibitemShut {NoStop}%
\end{thebibliography}
\end{document}